\documentclass{ptephy}

\usepackage{amsmath} 
\usepackage{amsthm} 
\usepackage{cite} 
\usepackage{graphicx} 
\usepackage{algorithmic} 
\usepackage{subfig} 
\usepackage{url} 
\usepackage{latexsym}
\usepackage{natbib}
\bibliography{sample}

\begin{document}
\title{Cosmological perturbations in the $(1+3+6)$-dimensional space-times}  
\author{\name{K. Tomita}{\ast}}
\address{\affil{}{Yukawa Institute for Theoretical Physics, Kyoto University, Kyoto 606-8502, Japan}
\rm{\email{ketomita@ybb.ne.jp}}}

\begin{abstract}
Cosmological perturbations in the $(1+3+6)$-dimensional space-times including
photon gas without viscous processes are studied on the basis of Abbott et al.'s
 formalism. Space-times consist
 of the outer space (the $3$-dimensional expanding section) and the inner space (the
$6$-dimensional section). The inner space expands initially and contracts later.
Abbott et al. derived only power-type solutions in the small wave-number limit 
which appear at the final stage of the space-times.
In this paper, we derive not only small wave-number solutions, but also large wave-number 
solutions. It is found that the latter solutions depend on the two wave-numbers $k_r$ and $k_R$
 (which are defined in the outer and inner spaces, respectively), and that the
 $k_r$-dependent and $k_R$-dependent parts dominate the total perturbations 
 when $(k_r/r(t))/(k_R/R(t)) \gg 1$ or $\ll 1$,  respectively, where $r(t)$ and $R(t)$ 
are the scale-factors in the outer 
and inner spaces. By comparing the behaviors of these perturbations, moreover,
 changes in the spectrum of perturbations in the outer space with time are discussed.
\end{abstract}

\maketitle

\section{Introduction}
From the viewpoint of analyzing cosmological perturbations and discussing the evolution
of their spectrum, we study the cosmological evolution of 
the $(1+3+6)$-dimensional space-times, in which it is assumed that our universe 
appears as an isotropic and homogeneous $10$-dimensional space-time and evolves to 
the state consisting of the $3$-dimensional inflating outer space and the $6$-dimensional
 collapsing outer space. This scenario is supported by the present super-string theory 
 (Kim et al. \citep{kim1,kim2} in a matrix model).
 
 In a previous paper\citep{tom}, we discussed the entropy production at the stage 
 when the above inflation and collapse coexist, and showed how viscous processes help 
 the increase of cosmological entropy; we also discussed the possibility that we satisfy, 
 at the same time,
 the condition that the entropy in the Guth level\citep{guth} is obtained and the 
 condition that the inner space decouples from the outer space.
 
 In this paper we study the evolution of cosmological perturbations in these space-times
 (in the case with no viscous processes),
 on the basis of Abbott et al.'s formalism\citep{abb}. They extended Bardeen's 
 gauge-invariant formalism\citep{bard} in the $4$-dimensional cosmological models to that 
 in the multi-dimensional models. 
 Abbott et al. derived only power-type solutions in the small wave-number limit 
which appear at the final stage of the space-times.
In this paper, we derive not only small wave-number solutions, but also large wave-number 
solutions. It is found that the latter solutions depend on the two wave-numbers $k_r$ and 
$k_R$ which are defined in the outer and inner spaces, respectively, and that $k_r$-dependent 
and $k_R$-dependent parts dominate the total perturbations when $(k_r/r(t))/(k_R/R(t)) \gg 1$ 
or $\ll 1$,  respectively, where $r(t)$ and $R(t)$ are the scale-factors in the outer 
and inner spaces. Using these solutions, we discuss the evolution of the $k_r$-dependence
 (spectrum) of perturbations in the outer space.
 
 In Sect. 2, we review our formalism based on that of Abbott et al., in which 
 the perturbed quantities and Einstein equations are shown and they are classified
 into three modes, i.e., scalar, vector, and tensor modes. A new equation to be solved 
in the scalar mode is introduced.
 In Sect. 3, we derive solutions for the perturbed equations in the scalar mode, and 
 approximate solutions in the cases of  $(k_r/r(t))/(k_R/R(t)) \gg 1$ 
or $\ll 1$ are shown. In Sects. 4 and 5, we derive solutions in the vector
and tensor modes, respectively.  Similarly approximate solutions in the cases of  
$(k_r/r(t))/(k_R/R(t)) \gg 1$ or $\ll 1$are shown. 
In Sect. 6, changes in the spectrum of perturbations with time are discussed. 
In Sect. 7, concluding remarks are given.  In Appendix A, we show the formulas of
harmonics and gauge transformations in outer and inner spaces. In Appendices B and C, 
we show the derivations of approximate solutions in the scalar mode in the cases of
$x (\equiv k_r \tau^{4/3}) \ll 1$ and  $y (\equiv k_R \tau^{2/3}) \ll 1$, respectively, 
corresponding to the above cases,
where $\tau \equiv t_0 - t$ and $t_0$ denotes the final time corresponding to 
$r \rightarrow \infty$ and $R = 0$.

\section{Formalism of the perturbation theory}
The background space-time is expressed in the form of a product of two homogeneous spaces $\rm{M_d}$ and $\rm{M_D}$ as
\begin{equation}
  \label{eq:a1}
ds^2 = -dt^2 + r^2(t)\ {}^d g_{ij} (x^k) \ dx^i dx^j + R^2(t)\ {}^Dg_{ab} (X^c) \ dX^a dX^b,
\end{equation}
where ${}^dg_{ij}$ and ${}^Dg_{ab}$ are the metrics of the outer space $\rm{M_d}$ and the 
inner space $\rm{M_D}$ with constant curvatures $K_r$ and $K_R$, respectively. Here the 
dimensions of $\rm{M_d}$ and $\rm{M_D}$ are $d = 3$ and $D = 6$. 
The inner space $\rm{M_D}$ expands initially and collapses after the maximum expansion with
$K_R = 1$, while the outer space  $\rm{M_d}$ continues to expand with $K_r = 0$ or $-1$. 
As the collapse and expansion in these spaces proceed, however, the curvature terms of
both spaces are negligible and the curvatures can be regarded  approximately as $K_r = K_R = 0$. Then the background metric is 
\begin{equation}
  \label{eq:a2}
\begin{split}
g_{00} &= -1, \quad g_{01} = g_{0a} = g_{ia} = 0, \\
g_{ij} &= r^2\ {}^d g_{ij}, \quad g_{ab} = R^2\ {}^D g_{ab}, 
\end{split}
\end{equation}
and the Ricci tensor is
\begin{equation}
  \label{eq:a3}
\begin{split}
R^0_0 &= - \left(d \frac{\ddot{r}}{r} + D \frac{\ddot{R}}{R}\right), \\
R^i_j &= -\delta^i_j \ \left[\left(\frac{\dot{r}}{r}\right)^. + \frac{\dot{r}}{r} \left(d \frac{\dot{r}}{r} +
D \frac{\dot{R}}{R}\right) + (d-1) \frac{K_r}{r^2}\right], \\
R^a_b &= -\delta^a_b \ \left[\left(\frac{\dot{R}}{R}\right)^. + \frac{\dot{R}}{R} \left(d \frac{\dot{r}}{r} +
D \frac{\dot{R}}{R}\right) + (D-1) \frac{K_R}{R^2}\right],
\end{split}
\end{equation}
where $i, j = 1, ..., d, \ a, b = d+1, ..., d+D$, and an overdot denotes $d/dt$. The background 
energy-momentum tensor is
\begin{equation}
  \label{eq:a4}
T^\mu_\nu = p \delta^\mu_\nu + (\rho + p) u^\mu u_\nu,
\end{equation}
where $u^\mu$ is the fluid velocity, $\rho$ the energy density, and $p$ the pressure. Here 
$\rho$ and $p$ are the common photon density and pressure in both spaces. 
The fluid is extremely hot and satisfies the equation of state $p = \rho/n$ of photon
gas, where $n = d + D = 9$. Einstein equations  are expressed as
\begin{equation}
  \label{eq:a5}
R^\mu_\nu = - 8\pi \bar{G} (T^\mu_\nu - \frac{1}{2} \delta^\mu_\nu T^\lambda_\lambda),
\end{equation}
where $\bar{G}$ is the $(1+d+D)$-dimensional gravitational constant. In the following, we 
set $8\pi \bar{G} = 1$. The background equation of motion for the matter is
\begin{equation}
  \label{eq:a6}
\frac{\dot{\rho}}{\rho+p} + d \frac{\dot{r}}{r} +D \frac{\dot{R}}{R} = 0.
\end{equation}

At the early stage, the expansion of the total universe is nearly isotropic (i.e. $r \propto R$). 
At the later stage, the inner space collapses after the maximum expansion, and at the final stage 
we have an approximate solution
\begin{equation}
  \label{eq:a7}
r = r_0 \ \tau^\eta,  \quad  R = R_0 \ \tau^\gamma  \quad (r_0, \ R_0 : const)
\end{equation}
with%
\begin{equation}
  \label{eq:a8}
\eta = \{1 - [D(n-1)/d]^{1/2} \}/n , \quad \gamma =  \{1 + [d(n-1)/D]^{1/2} \}/n,
\end{equation}
and $\tau = t_0 - t$, where $t_0$ is the final time corresponding to $R = 0$. For $d = 3$ and $D= 6$, we have
\begin{equation}
  \label{eq:a9}
\gamma = - \eta = 1/3.
\end{equation}
For the solutions (\ref{eq:a7}), Eqs. (\ref{eq:a4}) and (\ref{eq:a5}) lead to $R^0_0 = 0$ and 
$T^0_0 - \frac{1}{2} T^\mu_\mu \propto \rho$, so that we have 
\begin{equation}
  \label{eq:a10}
\rho = 0
\end{equation}
at the final stage.

\subsection{Classification of perturbations}

The simplest treatment of perturbations of geometrical and fluidal quantities is to expand 
them using harmonics, and to find the gauge-invariant quantities, as in Bardeen's theory for
 perturbations in the four-dimensional universe\citep{bard}. In the multi-dimensional universe
  consisting of 
the outer and inner homogeneous spaces $\rm{M_d}$ and $\rm{M_D}$ with different geometrical
 structures, we can have no harmonics in the $(d + D)$-dimensional space. 
Abbott et al.\citep{abb}  considered separate expansions in $\rm{M_d}$ and $\rm{M_D}$
using the harmonics defined in the individual spaces,  classified the perturbations in $\rm{M_d}$ 
and $\rm{M_D}$ individually as scalar (S), vector (V), and tensor (T), and classified the 
6 types of perturbations in $\rm{M_d} + \rm{M_D}$ as SS, SV, VS, VV, ST, and TS. 
The left and right sides of signatures correspond to the perturbations in $\rm{M_d}$ and 
 $\rm{M_D}$, respectively. The Helmholtz equations defining the harmonics are shown in 
 Appendix A.
 
 In the classification adopted by Abbott et al., the six types of perturbations are divided into 
 three groups:
 
 \noindent 1. scalar mode (SS),
 
 \noindent 2. vector mode (SV, VS, VV) , 

 \noindent 3. tensor mode (ST, TS).
 
 \noindent In this paper we call these three groups as ``modes'', corresponding to Abbott et al's 
 ``problems''.
 
 \subsection{Perturbed quantities}
 \subsubsection{The scalar mode}
 The metric perturbations are expressed as
\begin{equation}
  \label{eq:a11}
\begin{split}
g_{00} &= - (1 + 2 A \ q^{(0)} Q^{(0)}),\\
g_{0i} &= - r b^{(0)} q_i^{(0)} Q^{(0)},\quad g_{0a} = - R B^{(0)} q^{(0)} Q_a^{(0)},\\
g_{ij} &= r^2 [(1+2h_L q^{(0)} Q^{(0)})\ {}^dg_{ij} + 2h_T^{(0)}q_{ij}^{(0)} Q^{(0)}],\\
g_{ab} &= R^2 [(1+2H_L q^{(0)} Q^{(0)})\ {}^Dg_{ab} + 2H_T^{(0)}q^{(0)} Q_{ab}^{(0)}],\\
g_{ia} &= 2rR G^{(0)} q_i^{(0)} Q_a^{(0)},
\end{split}
\end{equation}
where $q^{(0)}, q_i^{(0)}, q_{ij}^{(0)}$ and  $Q^{(0)}, Q_a^{(0)}, Q_{ab}^{(0)}$ are scalar harmonics in  $\rm{M_d}$ and $\rm{M_D}$, respectively, and $A, b^{(0)}, B^{(0)}, h_L, H_L, h_T^{(0)}, H_T^{(0)}, $
 and $G^{(0)}$ are functions of $t$.

 The perturbations of fluid velocities and the energy-momentum tensor are expressed as  
\begin{equation}
  \label{eq:a12}
u^0 = 1 - A q^{(0)} Q^{(0)},\quad
u^i = \frac{v^{(0)}}{r} q^{(0)i} Q^{(0)},\quad
u^a = \frac{V^{(0)}}{R}  q^{(0)} Q^{(0)a},
\end{equation}
and
\begin{equation}
  \label{eq:a13}
\begin{split}
T^0_0 &= - \rho ( 1+ \delta \ q^{(0)} Q^{(0)}),\\
T^0_i &= r (\rho + p)(v^{(0)} - b^{(0)}) q_i^{(0)} Q^{(0)},\quad
T^0_a = R  (\rho + p)(V^{(0)} - B^{(0)}) q^{(0)} Q_a^{(0)},\\
T^i_j &= p (1 +\pi_L q^{(0)} Q^{(0)}) \delta^i_j,\quad
T^a_b = p (1 +\Pi_L q^{(0)} Q^{(0)}) \delta^a_b,\\
T^i_a &= 0,
\end{split}
\end{equation}
where we consider a perfect fluid, so that the anisotropic pressure terms
vanish and we have
\begin{equation}
  \label{eq:a14}
\pi_L  = \Pi_L = \delta.
\end{equation}

The metric perturbations in Eq. (\ref{eq:a11}) transform as shown in Appendix A for 
changes in the coordinates, and the following gauge-invariant quantities  are defined:
\begin{equation}
  \label{eq:a15}
\begin{split}
\Phi_h &= h_L + \frac{h_T^{(0)}}{d} + \frac{r}{k_r^{(0)}} \frac{\dot{r}}{r} b^{(0)} - 
\frac{r^2}{k_r^{(0)2}}\frac{\dot{r}}{r} \dot{h}_T^{(0)},\\
\Phi_H &= h_L + \frac{H_T^{(0)}}{D} + \frac{R}{k_R^{(0)}} \frac{\dot{R}}{R} B^{(0)} - 
\frac{R^2}{k_R^{(0)2}}\frac{\dot{R}}{R}\dot{H}_T^{(0)},
\end{split}
\end{equation}
\begin{equation}
  \label{eq:a16}
\begin{split}
\Phi_A^{(r)} &= A + \frac{r}{k_r^{(0)}}\dot{b}^{(0)} + \frac{r}{k_r^{(0)}} \left(\frac{\dot{r}}{r} +
D\frac{\dot{R}}{R}\right) b^{(0)} \\
&- \frac{r^2}{k_r^{(0)2}}\left[\ddot{h}_T^{(0)} + \left(2\frac{\dot{r}}{r} +
D\frac{\dot{R}}{R}\right) \dot{h}_T^{(0)}\right] + D\left(H_L +  \frac{H_T^{(0)}}{D}\right),\\
\Phi_A^{(R)} &= A + \frac{R}{k_R^{(0)}} \dot{B}^{(0)} + \frac{R}{k_R^{(0)}} \left(d\frac{\dot{r}}{r} +
\frac{\dot{R}}{R}\right) B^{(0)} \\
&- \frac{R^2}{k_R^{(0)2}} \left[\ddot{H}_T^{(0)} + \left(d\frac{\dot{r}}{r} +
2\frac{\dot{R}}{R}\right) \dot{H}_T^{(0)}\right] + d\left(h_L +  \frac{h_T^{(0)}}{d}\right).
\end{split}
\end{equation}
The gauge-invariant quantities $\Phi_h$ and $\Phi_A^{(r)}$ in the outer space 
correspond to the gauge-invariant perturbations defined by Bardeen\citep{bard} in the
$(1+3)$-dimensional usual universes, and  $\Phi_H$ and $\Phi_A^{(R)}$ in the inner 
space are similar to the above quantities. $\Phi_h$ and  $\Phi_H$ represent the 
curvature perturbations in both spaces.

The gauge-invariant quantities for fluid velocity and energy density perturbations are given by
\begin{equation}
  \label{eq:a17}
v_s^{(0)} = v^{(0)} - \frac{r}{k_r^{(0)}} \dot{h}_T^{(0)}, \quad
V_s^{(0)} = V^{(0)} - \frac{R}{k_R^{(0)}} \dot{H}_T^{(0)}, 
\end{equation}
and
\begin{equation}
  \label{eq:a18}
\begin{split}
\epsilon_m &= \delta +\frac{n+1}{n}\left[d\frac{\dot{r}}{k_r^{(0)}} (v^{(0)}-b^{(0)})+
D\frac{\dot{R}}{k_R^{(0)}} (V^{(0)}-B^{(0)})\right] , \\
\epsilon_g &=  \delta -\frac{n+1}{n}\left[d\frac{\dot{r}}{k_r^{(0)}} (b^{(0)} - \frac{r}
{k_r^{(0)}}\dot{h}_T^{(0)}) + D\frac{\dot{R}}{k_R^{(0)}} (B^{(0)} - \frac{R}
{k_R^{(0)}}\dot{H}_T^{(0)})\right] ,
\end{split}
\end{equation}
As a gauge-invariant quantity that has no counterpart in the usual universe, we have
\begin{equation}
  \label{eq:a19}
\Phi_G = G^{(0)} - \frac{1}{2} \frac{k_R^{(0)}}{k_r^{(0)}}\frac{r}{R} h_T^{(0)} - 
 \frac{1}{2} \frac{k_r^{(0)}}{k_R^{(0)}}\frac{R}{r} H_T^{(0)},
\end{equation}
which was introduced by Abbott et al.\citep{abb}.

\subsubsection{The vector mode}
The metric perturbations are 
\begin{equation}
  \label{eq:a20}
\begin{split}
g_{00} &= - 1,\\ 
g_{0i} &= - r b^{(1)} q_i^{(1)} Q^{(0)},\quad
g_{0a} = - R B^{(1)} q^{(0)} Q_a^{(1)},\\
g_{ij} &= r^2 ({}^d g_{ij} + 2h_T^{(1)}q_{ij}^{(1)} Q^{(0)}),\quad
g_{ab} = R^2  ({}^D g_{ab} + 2H_T^{(1)}q^{(0)} Q_{ab}^{(1)}),\\
g_{ia} &= 2rR (G^{(01)} q_i^{(0)} Q_a^{(1)} +G^{(10)} q_i^{(1)} Q_a^{(0)} +G^{(11)} q_i^{(1)} Q_a^{(1)}).
\end{split}
\end{equation}
The perturbed fluid velocity is
\begin{equation}
  \label{eq:a21}
\begin{split}
u^0 &= 1 ,\\
u^i &= \frac{v^{(1)}}{r} q^{(1)i} Q^{(0)},\quad u^a = \frac{V^{(1)}}{R}  q^{(0)} Q^{(1)a},
\end{split}
\end{equation}
and the perturbed energy-momentum tensor is
\begin{equation}
  \label{eq:a22}
\begin{split}
T^0_0 &= - \rho,\\
T^0_i &= r (\rho + p)(v^{(1)} - b^{(1)}) q_i^{(1)} Q^{(0)},\\
T^0_a &= R  (\rho + p)(V^{(1)} - B^{(1)}) q^{(0)} Q_a^{(1)},\\
T^i_j &= p \delta^i_j,\quad T^a_b = p \delta^a_b,\quad T^i_a = 0,
\end{split}
\end{equation}
where we neglected anisotropic stresses.

For the VS part of the vector mode, we have the gauge-invariant metric perturbations 
defined by
\begin{equation}
  \label{eq:a23}
\Psi_r = b^{(1)} - \frac{r}{k_r^{(1)}} \dot{h}_T^{(1)} ,\quad
J_r  = G^{(10)} - \frac{1}{2} \frac{k_R^{(0)}}{k_r^{(1)}} \frac{r}{R} h_T^{(1)},
\end{equation}
and fluidal perturbations are
\begin{equation}
  \label{eq:a24}
v_s^{(1)} = v^{(1)} - \frac{r}{k_r^{(1)}} \dot{h}_T^{(1)} ,\quad
v_c  = v^{(1)} - b^{(1)} = v_s^{(1)}  -\Psi_r .
\end{equation}
For the SV part of the vector mode,
\begin{equation}
  \label{eq:a25}
\Psi_R = B^{(1)} - \frac{R}{k_R^{(1)}} \dot{H}_T^{(1)} ,\quad
J_R  = G^{(01)} - \frac{1}{2} \frac{k_r^{(0)}}{k_R^{(1)}} \frac{R}{r} H_T^{(1)},
\end{equation}
and fluidal perturbations are
\begin{equation}
  \label{eq:a26}
V_s^{(1)} = V^{(1)} - \frac{R}{k_R^{(1)}} \dot{H}_T^{(1)} ,\quad
V_c  = V^{(1)} - B^{(1)} = V_s^{(1)}  -\Psi_R .
\end{equation}
For the VV part, we have only one gauge-invariant quantity $G^{(11)}$.

\subsubsection{The tensor mode}
We have only metric perturbations given by
\begin{equation}
  \label{eq:a27}
\begin{split}
g_{00} &= - 1, \quad g_{0i} = g_{0a} = g_{ia} = 0, \\ 
g_{ij} &= r^2 ({}^d g_{ij} + 2h_T^{(2)}q_{ij}^{(2)} Q^{(0)}),\\
g_{ab} &= R^2  ({}^D g_{ab} + 2H_T^{(2)}q^{(0)} Q_{ab}^{(2)}),
\end{split}
\end{equation}
and have no fluidal perturbations, where have we neglected anisotropic stresses.
In this mode, $h_T^{(2)}$ and $H_T^{(2)}$ correspond to the TS and ST parts of curvature
 perturbations and they themselves are gauge-invariant.

\subsection{Perturbed Einstein equations}
The perturbed Einstein equations are
\begin{equation}
  \label{eq:a28}
\delta G^\mu_\nu \equiv \delta R^\mu_\nu - \frac{1}{2} \delta^\mu_\nu \delta 
R^\lambda_\lambda = - 8\pi \bar{G} \delta T^\mu_\nu.
\end{equation}

\subsubsection{The scalar mode}
First we take up the following three relations which hold in the perturbed Einstein 
equations 
\begin{equation}
  \label{eq:a29}
\begin{split}
\delta G^i_j &- \frac{1}{d} \delta^i_j  \delta G^k_k = 0, \quad
\delta G^a_b - \frac{1}{D} \delta^a_b \delta G^c_c = 0, \\ 
\delta G^i_a &= 0.
\end{split}
\end{equation}
Using the expressions of $\delta R^\mu_\nu$ given in the Appendix of Ref.\citep{abb},
 we obtain three relations between the gauge-invariant quantities from
the above relations:
\begin{equation}
  \label{eq:a30}
\Phi_A^{(r)} + (d-2) \Phi_h + 2(k_R^{(0)}/R)^2 \tilde{\Phi}_G = 0,
\end{equation}
\begin{equation}
  \label{eq:a31}
\Phi_A^{(R)} + (D-2) \Phi_H + 2(k_r^{(0)}/r)^2 \tilde{\Phi}_G = 0,
\end{equation} 
\begin{equation}
  \label{eq:a32}
[(k_r^{(0)}/r)^2 + (k_R^{(0)}/R)^2] \tilde{\Phi}_G = \ - \ \Xi_G, 
\end{equation}
where 
\begin{equation}
  \label{eq:a33}
\tilde{\Phi}_G \equiv \frac{rR}{k_r^{(0)} k_R^{(0)}} \Phi_G, 
\end{equation}
$\Phi_G$ is a gauge-invariant quantity defined in Eq.(\ref{eq:a19}),  and
\begin{equation}
  \begin{split}
    \label{eq:a34a}
\Xi_G &\equiv  \frac{rR}{k_r^{(0)} k_R^{(0)}} \left\{ \ddot{\Phi}_G +\left(d\frac{\dot{r}}{r}+
D\frac{\dot{R}}{R}\right) \dot{\Phi}_G 
-  \left[(d-1)\frac{K_r}{r^2}+(D-1)\frac{K_R}{R^2}+\left(\frac{\dot{r}}{r}
-\frac{\dot{R}}{R}\right)^2 \right]  \Phi_G \right\} \\
&- \left(\frac{\dot{r}}{r} -\frac{\dot{R}}{R}\right)\left(\frac{r}{\dot{r}}\Phi_h - \frac{R}{\dot{R}}\Phi_H
 - \Phi_6\right).
\end{split}
\end{equation}
This equation is rewritten in terms of $\tilde{\Phi}_G$ as
\begin{equation}
  \label{eq:a34}
  \begin{split}
\Xi_G &= \ddot{\tilde{\Phi}}_G +\left(d\frac{\dot{r}}{r}+
D\frac{\dot{R}}{R}\right) \dot{\tilde{\Phi}}_G 
-  \left[(d-1)\frac{K_r}{r^2}+(D-1)\frac{K_R}{R^2}+\left(\frac{\dot{r}}{r}
-\frac{\dot{R}}{R}\right)^2 \right]  \tilde{\Phi}_G  \\
&-\left(\frac{\dot{r}}{r} -\frac{\dot{R}}{R}\right)\left(\frac{r}{\dot{r}}\Phi_h - \frac{R}{\dot{R}}\Phi_H
 - \Phi_6\right),
\end{split}
\end{equation}
 where we have used the relation $rR =$ const. 
 
Here $\Phi_6$ is an auxiliary gauge-invariant quantity defined by
\begin{equation}
  \label{eq:a35}
\frac{\dot{r}}{r} \frac{\dot{R}}{R} \Phi_6 \equiv \frac{\dot{R}}{R} \left(h_L +\frac{h_T^{(0)}}{d}\right) -
\frac{\dot{r}}{r} \left(H_L +\frac{H_T^{(0)}}{D}\right),  
\end{equation} 
which satisfies 
\begin{equation}
  \label{eq:a36}
    \begin{split}
\dot{\Phi}_6 +  \left(d\frac{\dot{r}}{r}+D\frac{\dot{R}}{R}\right) \Phi_6 &= 
 \left(\frac{r}{\dot{r}}\right)^2 \left\{\frac{\dot{r}}{r}\dot{\Phi}_h - \left[\left(\frac{\dot{r}}{r}\right)^. 
 - D\frac{\dot{r}}{r} \frac{\dot{R}}{R}\right] \Phi_h\right\} - \Phi_A^{(r)} \\
 &-  \left(\frac{R}{\dot{R}}\right)^2\left\{\frac{\dot{R}}{R}\dot{\Phi}_H - \left[\left(\frac{\dot{R}}{R}\right)^. - 
 d\frac{\dot{r}}{r} \frac{\dot{R}}{R}\right]\Phi_H\right\} + \Phi_A^{(R)}.
\end{split}
\end{equation} 

Next from another relation
\begin{equation}
  \label{eq:a37}
\delta G^0_0 -\frac{d}{k_r^{(0)2}} \frac{\dot{r}}{r}\delta G_i^{0|i} - \frac{D}{k_R^{(0)2}} 
\frac{\dot{R}}{R}\delta G_a^{0|a} = -\rho \epsilon_m q^{(0)} Q^{(0)},
\end{equation} 
we obtain 
\begin{equation}
  \label{eq:a38}
(d-1)\left(\frac{k_r^{(0)}}{r}\right)^2 \left(1- \frac{dK_r}{k_r^{(0)2}}\right)
\Phi_h  + (D-1)\left(\frac{k_R^{(0)}}{R}\right)^2 \left(1- \frac{DK_R}{k_R^{(0)2}}\right)\Phi_H 
+ \Xi_\epsilon =  -\rho \epsilon_m ,
\end{equation} 
where
\begin{equation}
  \label{eq:a39}
    \begin{split}
\Xi_\epsilon &\equiv \left[d \frac{\dot{r}}{r} \left(\frac{k_R^{(0)}}{R}\right)^2 + D \frac{\dot{R}}{R}
\left(\frac{k_r^{(0)}}{r}\right)^2\right] \left[\frac{1}{2} \frac{r}{\dot{r}}\Phi_h + \frac{1}{2} 
\frac{R}{\dot{R}}\Phi_H -(\tilde{\Phi}_G)^. \right] \\
&+ 2 \tilde{\Phi}_G \left[\left(\frac{k_r^{(0)} k_R^{(0)}}{rR}\right)^2 + d\left(\frac{\dot{r}}{r} 
\frac{k_R^{(0)}}{R}\right)^2 + D\left(\frac{\dot{R}}{R} \frac{k_r^{(0)}}{r}\right)^2\right] \\
&+ \frac{1}{2} \Phi_6 \left\{d\frac{\dot{r}}{r} \left[\left(\frac{k_R^{(0)}}{R}\right)^2 + 
2D\left(\frac{\dot{R}}{R}\right)^2\right]
- D\frac{\dot{R}}{R} \left[\left(\frac{k_r^{(0)}}{r}\right)^2 + 2d\left(\frac{\dot{r}}{r}\right)^2\right]\right\}.
\end{split}
\end{equation} 

As one of equations describing the time development of $\Phi_h$ and $\Phi_H$,  we have  
\begin{equation}
  \label{eq:a40}
\delta G^\lambda_\lambda = 0,
\end{equation} 
which is expressed using gauge-invariant quantities as
\begin{equation}
  \label{eq:a41}
    \begin{split}
d \{\ddot{\Phi}_h &+ \left[(d+1)\frac{\dot{r}}{r}+ 2D\frac{\dot{R}}{R}\right] \dot{\Phi}_h - 
\frac{r}{\dot{r}} [\left(\frac{\dot{r}}{r}\right)^{..} +(d+1)\frac{\dot{r}}{r}\left(\frac{\dot{r}}{r}\right)^. 
- D(d+1)\left(\frac{\dot{r}}{r}\right)^2 \frac{\dot{R}}{R} \\
&-D^2\frac{\dot{r}}{r} \left(\frac{\dot{R}}{R}\right)^2 - D\frac{\dot{r}}{r}\left(\frac{\dot{R}}{R}\right)^.
 + \frac{d-1}{d}\frac{\dot{r}}{r}\left(\frac{k_r^{(0)}}{r}\right)^2] \Phi_h \\
& -\frac{\dot{r}}{r}\dot{\Phi}_A^{(r)} - \left[2\left(\frac{\dot{r}}{r}\right)^. + (d+1) \left(\frac{\dot{r}}{r}\right)^2 
+ D\frac{\dot{r}}{r}\frac{\dot{R}}{R} + \frac{k_r^{(0)2}}{dr^2}\right] \Phi_A^{(r)} \} \\
&+ D \{l.c. \leftrightarrow u.c.\} + \frac{1}{2}\Phi_6 \left[d\left(\frac{\ddot{r}}{r}\right)^. 
- D\left(\frac{\ddot{R}}{R}\right)^. - \rho \frac{n+1}{n} \left(d \frac{\dot{r}}{r}- D\frac{\dot{R}}{R}\right)
\right] = 0,
\end{split}
\end{equation} 
where $\{l.c. \leftrightarrow u.c.\}$ means the terms (in $\{ \ \}$) given by the exchanges 
$r \leftrightarrow R$ and $d \leftrightarrow D$.

As another equation describing the time development of $\Phi_h$ and $\Phi_H$,  we adopt
\begin{equation}
  \label{eq:a42}
\frac{1}{d} \delta G^i_i - \frac{1}{D} \delta G^a_a = 0,
\end{equation} 
which holds because $\delta T^i_j = p \delta\ q^{(0)} Q^{(0)} \delta ^i_j$ and  
 $\delta T^a_b= p \delta\ q^{(0)} Q^{(0)} \delta ^a_b$.  \ Equation (\ref{eq:a42}) is expressed 
 using the gauge-invariant quantities as 
\begin{equation}
  \label{eq:a43}
    \begin{split}
\langle \ddot{\Phi}_h &+ \left[2d \frac{\dot{r}}{r}+ (D-2d)\frac{\dot{R}}{R}\right] \dot{\Phi}_h +
\{\left(\frac{k_R^{(0)}}{R}\right)^2-2d\left[\left(\frac{\dot{R}}{R}\right)^. +\frac{\dot{R}}{R}\left(d\frac{\dot{r}}{r} +
D\frac{\dot{R}}{R}\right)\right] \\
&+ \frac{2(d-1)}{d}\left(\frac{k_r^{(0)}}{r}\right)^2 \left(1- \frac{dK_r}{k_r^{(0)2}}\right) \} \Phi_h 
+\left\{ \frac{k_r^{(0)2}}{dr^2}- 2\left[\left(\frac{\dot{r}}{r}\right)^. + \frac{\dot{r}}{r}\left(d\frac{\dot{r}}{r} +
D\frac{\dot{R}}{R}\right)\right] \right\} \Phi_A^{(r)} \\
&- \frac{\dot{r}}{r} \dot{\Phi}_A^{(r)} \rangle 
- \langle l.c. \leftrightarrow u.c. \rangle +\frac{2k_r^{(0)} k_R^{(0)}}{rR} \left(\frac{1}{d} - \frac{1}{D}\right) \Phi_G
+ \Xi_7 \Phi_7 + 2n  \frac{\dot{r}}{r} \frac{\dot{R}}{R} \dot{\Phi}_7 = 0,
\end{split}
\end{equation} 
where $\langle l.c. \leftrightarrow u.c.\rangle$ means the terms (in $\langle \ \rangle$) given by the exchanges 
$r \leftrightarrow R$ and $d \leftrightarrow D$,
\begin{equation}
  \label{eq:a44}
    \begin{split}
\Xi_7 &\equiv \{\frac{d(d-1)}{n} \frac{\dot{r}}{r} \left[\frac{K_r}{r^2} -\frac{\ddot{r}}{r}
+ \left(\frac{\dot{r}}{r}\right)^2\right] + \left(\frac{k_R^{(0)}}{r}\right)^2 \frac{\dot{r}}{r} +\left(2n - \frac{dD}{n}\right) 
\frac{\dot{R}}{R} \frac{\ddot{r}}{r} \\
&+ \left[2n(d-1) + \frac{Dd}{n}\right] \left(\frac{\dot{r}}{r}\right)^2 \frac{\dot{R}}{R}\} + \{l.c. \leftrightarrow u.c.\}
\end{split}
\end{equation} 
and $\Phi_7$ is another auxiliary gauge-invariant quantity defined by
\begin{equation}
  \label{eq:a45}
\Phi_7 \equiv \frac{r}{k_r^{(0)}}b^{(0)} -\frac{R}{k_R^{(0)}}B^{(0)} - \left(\frac{r}{k_r^{(0)}}\right)^2 \dot{h}_T^{(0)} 
+ \left(\frac{R}{k_R^{(0)}}\right)^2 \dot{H}_T^{(0)} ,
\end{equation} 
satisfying the relation
\begin{equation}
  \label{eq:a46}
 \frac{\dot{R}}{R} \Phi_h -  \frac{\dot{r}}{r} \Phi_H = \frac{\dot{r}}{r}  \frac{\dot{R}}{R} 
 (\Phi_6 + \Phi_7).
\end{equation} 

In Ref.\citep{abb}, Eq. (\ref{eq:a42}) was not adopted as the equation 
to be solved, but it is a fundamental equation to be solved to derive $\Phi_h$ and
$\Phi_H$ in general situations. They paid attention only to the case when 
$\rho \rightarrow 0$ at the final stage  $r 
\rightarrow \infty$ and $R \rightarrow 0$. In this case,  Eq.(\ref{eq:a38}) with $\rho 
\epsilon = 0$ may be one of the conditions for constraining the behaviors of $\Phi_h$ and 
$\Phi_H$, and they could derive the behavior of  $\Phi_h$ and $\Phi_H$ using it in the limit of 
small wave-numbers. In present paper, however, we use Eqs. (\ref{eq:a42}) and 
(\ref{eq:a43}) to derive their behaviors in more general cases including the case
of large wave-numbers. Then, equations to be solved are Eqs. (\ref{eq:a32}), 
(\ref{eq:a36}),  (\ref{eq:a41}), and  (\ref{eq:a43}) for the four quantities 
$ \tilde{\Phi}_G, \Phi_6, \Phi_h,$ and $\Phi_H$.

\subsubsection{The vector mode}
In the VS case, we have the following three equations from $\delta G^\mu_\nu = 
\delta T^\mu_\nu $
\begin{equation}
  \label{eq:a47}
 \frac{1}{2} \left\{\left(\frac{k_r^{(1)}}{r}\right)^2 \left[1-(d-1)\frac{K_r}{k_r^{(1)2}}\right] +
 \left(\frac{k_R^{(0)}}{R}\right)^2\right\} \Psi_r - \frac{k_R^{(0)}}{R}
 \left[\dot{J}_r - \left(\frac{\dot{r}}{r} - \frac{\dot{R}}{R}\right) J_r\right] = - (\rho +p) v_c, 
\end{equation} 
\begin{equation}
  \label{eq:a48}
 \dot{\Psi}_r + \left[(d-1) \frac{\dot{r}}{r} + D\frac{\dot{R}}{R}\right] \Psi_r
 + 2 \frac{k_R^{(0)}}{R} J_r= 0,
\end{equation} 
and
\begin{equation}
  \label{eq:a49}
      \begin{split}
  \frac{1}{2} \dot{\Psi}_r &+\frac{1}{2}\left[(d+1) \frac{\dot{r}}{r} + (D-2)
  \frac{\dot{R}}{R}\right] \Psi_r - \frac{R}{k_R^{(0)}} \{\ddot{J}_r + \left(d\frac{\dot{r}}{r}+
 D\frac{\dot{R}}{R}\right) \dot{J}_r\\
 &+ \left[\left(\frac{k_r^{(1)}}{r}\right)^2 - (D-1)\frac{K_R}{R^2} - \left(\frac{\dot{r}}{r}
 - \frac{\dot{R}}{R}\right)^2\right] J_r  \} = 0.      
       \end{split}
\end{equation} 

In the SV case, we have similarly
\begin{equation}
  \label{eq:a50}
 \frac{1}{2}  \left\{ \left(\frac{k_r^{(0)}}{r}\right)^2 +  \left(\frac{k_R^{(1)}}{R}\right)^2 
  \left[1-(D-1)\frac{K_R}{k_R^{(1)2}}\right] \right\} \Psi_R - \frac{k_r^{(0)}}{r}
  \left[\dot{J}_R +  \left(\frac{\dot{r}}{r} - \frac{\dot{R}}{R}\right) J_R\right] = - (\rho +p) V_c, 
\end{equation} 
\begin{equation}
  \label{eq:a51}
 \dot{\Psi}_R +  \left[d \frac{\dot{r}}{r} + (D-1)\frac{\dot{R}}{R}\right] \Psi_R
 + 2 \frac{k_r^{(0)}}{r} J_R = 0,
\end{equation} 
and
\begin{equation}
  \label{eq:a52}
        \begin{split}
 \frac{1}{2} \dot{\Psi}_R &+\frac{1}{2}  \left[(D+1) \frac{\dot{R}}{R} + 
 (d-2) \frac{\dot{r}}{r}\right] \Psi_R - \frac{r}{k_r^{(0)}} \{\ddot{J}_R +  \left(d\frac{\dot{r}}{r}+
 D\frac{\dot{R}}{R}\right) \dot{J}_R \\
 &+ \left[ \left(\frac{k_R^{(1)}}{R}\right)^2 - (d-1)\frac{K_r}{r^2} -  \left(\frac{\dot{r}}{r}
 - \frac{\dot{R}}{R}\right)^2\right] J_R \} = 0.
  \end{split}
\end{equation} 

 In the VV case, we have for $G^{(11)}$
\begin{equation}
  \label{eq:a53}
 \ddot{G}^{(11)} +  \left(d \frac{\dot{r}}{r} + D\frac{\dot{R}}{R}\right) \dot{G}^{(11)} + 
  \left[ \left(\frac{k_r^{(1)}}{r}\right)^2 + \left(\frac{k_R^{(1)}}{R}\right)^2 -  
  \left(\frac{\dot{r}}{r}- \frac{\dot{R}}{R}\right)^2\right]  G^{(11)} = 0.
\end{equation} 

\subsubsection{The tensor mode}     
In the TS case, we have for the gauge-invariant quantity $h_T^{(2)} $        
\begin{equation}
  \label{eq:a54}
 \ddot{h}_T^{(2)} +  \left(d \frac{\dot{r}}{r} + D\frac{\dot{R}}{R}\right) \dot{h}_T^{(2)} + 
 \left[ \left(\frac{k_r^{(2)}}{r}\right)^2 +  \left(\frac{k_R^{(0)}}{R}\right)^2  + 2 \frac{K_r}{r^2}\right] h_T^{(2)} =0,
\end{equation} 
and  in the ST case  for $H_T^{(2)} $  
\begin{equation}
  \label{eq:a55}
 \ddot{H}_T^{(2)} +  \left(d \frac{\dot{r}}{r} + D\frac{\dot{R}}{R}\right) \dot{H}_T^{(2)} + 
 \left[ \left(\frac{k_r^{(0)}}{r}\right)^2 +  \left(\frac{k_R^{(2)}}{R}\right)^2  + 2 \frac{K_R}{R^2}\right] H_T^{(2)} =0.
\end{equation} 
In these cases, we have $h_T^{(2)}$ and $H_T^{(2)}$ representing the tensor components, 
but no scalar and vector quantities. 


\section{Solutions in the scalar mode}
Let us derive various approximate solutions for the equations of perturbations, at the 
final stage in the model with $d = 3$ and $D = 6$. In this model we have the relation $r R =$
const (cf. Eqs.(\ref{eq:a7}) and (\ref{eq:a9})), which is useful to simplify the 
derivation of solutions.

\subsection{Basic equations}
In the present model, Eq. (\ref{eq:a32}) can be expressed as
\begin{equation}
  \label{eq:b1}
{\tilde{\Phi}_G}'' + \frac{1}{\tau} {\tilde{\Phi}_G}' + \left[ \left(\frac{k_r^{(0)}}{r}\right)^2 +
  \left(\frac{k_R^{(0)}}{R}\right)^2 -\frac{4/9}{\tau^2}\right] \tilde{\Phi}_G = 2\Phi_h + 2\Phi_H - 
 \frac{2}{3} \Phi_6/\tau,
\end{equation} 
where $\tau = t_0 - t$, $'$ denotes $\partial/\partial \tau$, and $\tilde{\Phi}_G$ is
defined by Eq. (\ref{eq:a33}). 
 
Next, eliminating $\Phi_A^{(r)}$ and  $\Phi_A^{(R)}$ in Eq. (\ref{eq:a36}) by the use of Eqs. 
(\ref{eq:a30}) and (\ref{eq:a31}), we express the equation for $\Phi_6$ as
\begin{equation}
  \label{eq:b2}
{\Phi_6}' + \frac{1}{\tau} \Phi_6 = 3\tau \left({\Phi_h}' + \frac{8/3}{\tau} \Phi_h +
{\Phi_H}' + \frac{4/3}{\tau} \Phi_H\right) + 2 \left[ \left(\frac{k_r^{(0)}}{r}\right)^2 - 
 \left(\frac{k_R^{(0)}}{R}\right)^2\right] \tilde{\Phi}_G,
\end{equation} 
Eliminating $\Phi_A^{(r)}$ and  $\Phi_A^{(R)}$, Eq.(\ref{eq:a41}) can be expressed as
\begin{equation}
  \label{eq:b3}
    \begin{split}
{\Phi_h}'' &+ \frac{7/3}{\tau}{\Phi_h}' +  \left[\frac{1}{3}  \left(\frac{k_r^{(0)}}{r}\right)^2 
- \frac{32/9}{\tau^2}\right] \Phi_h \\
&+ 2 \left\{{\Phi_H}'' +\frac{5/3}{\tau}{\Phi_H}' +  \left[\frac{1}{6} \left(\frac{k_R^{(0)}}{R}\right)^2 
- \frac{8/9}{\tau^2}\right] \Phi_H\right\} -  \frac{8}{9}\tau^{-3} \Phi_6 \\
 &+  \left[\frac{4}{3}  \left(\frac{k_r^{(0)}k_R^{(0)}}{rR}\right)^2 +
 \frac{4}{3\tau^2} \left(\frac{k_R^{(0)}}{R}\right)^2 \right]\tilde{\Phi}_G +\frac{2}{3\tau} 
  \left[2 \left(\frac{k_r^{(0)}}{r}\right)^2 -
  \left(\frac{k_R^{(0)}}{R}\right)^2\right] {\tilde{\Phi}_G}' = 0.
\end{split}
\end{equation} 
Equation (\ref{eq:a43}) can similarly be expressed as 
\begin{equation}
  \label{eq:b4}
    \begin{split}
{\Phi_h}'' &+ \frac{11/3}{\tau} {\Phi_h}' + 2  \left(\frac{k_R^{(0)}}{R}\right)^2  \Phi_h 
- {\Phi_H}'' -\frac{7/3}{\tau}{\Phi_H}' -2  \left(\frac{k_r^{(0)}}{r}\right)^2 \Phi_H \\
&-\frac{2}{\tau^2}  \left\{{\Phi_6}' - \frac{1}{\tau}  \left[1+\frac{1}{6}  \left(\frac{k_r^{(0)}\tau}{r}\right)^2 
- \frac{1}{6}  \left(\frac{k_R^{(0)}\tau}{R}\right)^2 \right]\Phi_6\right\} - \frac{2\gamma }{\tau}
 \left[  \left(\frac{k_r^{(0)}}{r}\right)^2 +   \left(\frac{k_R^{(0)}}{R}\right)^2\right] {\tilde{\Phi}_G}' \\
&- \frac{4/9}{\tau^2}  \left[  \left(\frac{k_r^{(0)}}{r}\right)^2 - \left(\frac{k_R^{(0)}}{R}\right)^2
\right] \tilde{\Phi}_G
= 0,
\end{split}
\end{equation} 
or furthermore eliminating ${\Phi_6}'$ by use of Eq. (\ref{eq:b2}), we obtain
\begin{equation}
  \label{eq:b5}
    \begin{split}
{\Phi_h}'' &- \frac{7/3}{\tau} {\Phi_h}' +  \left[2  \left(\frac{k_R^{(0)}}{R}\right)^2-\frac{16}{\tau^2}\right]  \Phi_h 
- {\Phi_H}'' -\frac{25/3}{\tau}{\Phi_H}' - \left[2  \left(\frac{k_r^{(0)}}{r}\right)^2 +\frac{8}{\tau^2}\right] \Phi_H \\
&+\frac{2}{\tau^3}  \left[2 +\frac{1}{6} \left(\frac{k_r^{(0)}\tau}{r}\right)^2 
- \frac{1}{6}  \left(\frac{k_R^{(0)}\tau}{R}\right)^2 \right]\Phi_6 - \frac{2/3}{\tau}
 \left[  \left(\frac{k_r^{(0)}}{r}\right)^2 +  \left(\frac{k_R^{(0)}}{R}\right)^2\right] {\tilde{\Phi}_G}' \\
&- \frac{40/9}{\tau^2}  \left[  \left(\frac{k_r^{(0)}}{r}\right)^2 -  \left(\frac{k_R^{(0)}}{R}\right)^2\right] \tilde{\Phi}_G
= 0.
\end{split}
\end{equation} 

Equations  (\ref{eq:b3}) and (\ref{eq:b5}) are rewritten as the equations giving only  ${\Phi_h}'' $ and ${\Phi_H}'' $
as follows :
\begin{equation}
  \label{eq:b6}
    \begin{split}
3{\Phi_h}'' &- \frac{7/3}{\tau}{\Phi_h}' +  \left[\frac{1}{3}  \left(\frac{k_r^{(0)}}{r}\right)^2 +
4 \left(\frac{k_R^{(0)}}{R}\right)^2-\frac{320/9}{\tau^2}\right] \Phi_h \\
&- \frac{40/3}{\tau}{\Phi_H}' -  \left[\frac{160/9}{\tau^2} +4 \left(\frac{k_r^{(0)}}{r}\right)^2 - 
\frac{1}{3} \left(\frac{k_R^{(0)}}{R}\right)^2\right] \Phi_H  +  \frac{4}{\tau^3}  \left[\frac{1}{6} \left(\frac{k_r^{(0)}\tau}
{r}\right)^2- \frac{1}{6} \left(\frac{k_R^{(0)}\tau}{r}\right)^2 + \frac{16}{9}\right] \Phi_6 \\
&+  \left\{\frac{4}{3}  \left(\frac{k_r^{(0)}k_R^{(0)}}{rR}\right)^2 +
 \frac{1}{9\tau^2} \left[92  \left(\frac{k_R^{(0)}}{R}\right)^2 - 80 \left(\frac{k_r^{(0)}}{r}\right)^2\right]\right\} \tilde{\Phi}_G 
 -\frac{2}{\tau}  \left(\frac{k_R^{(0)}}{R}\right)^2  {\tilde{\Phi}_G}' = 0,
\end{split}
\end{equation} 
and 
\begin{equation}
  \label{eq:b7}
    \begin{split}
3{\Phi_H}'' &+ \frac{35/3}{\tau}{\Phi_H}' +  \left[2  \left(\frac{k_r^{(0)}}{r}\right)^2 +\frac{1}{3} 
 \left(\frac{k_R^{(0)}}{R}\right)^2 +\frac{56/9}{\tau^2}\right] \Phi_H \\
&+ \frac{14/3}{\tau}{\Phi_h}' +  \left[\frac{112/9}{\tau^2} +\frac{1}{3} \left(\frac{k_r^{(0)}}{r}\right)^2 - 
2  \left(\frac{k_R^{(0)}}{R}\right)^2\right] \Phi_h  -  \frac{2}{\tau^3}  \left[\frac{1}{6} \left(\frac{k_r^{(0)}\tau}
{r}\right)^2- \frac{1}{6} \left(\frac{k_R^{(0)}\tau}{r}\right)^2 + \frac{22}{9}\right] \Phi_6 \\
&+ \left\{\frac{4}{3}  \left(\frac{k_r^{(0)}k_R^{(0)}}{rR}\right)^2 +
 \frac{1}{9\tau^2} \left[40  \left(\frac{k_r^{(0)}}{r}\right)^2 - 28 \left(\frac{k_R^{(0)}}{R}\right)^2\right]
 \right\} \tilde{\Phi}_G +\frac{2}{\tau}  \left(\frac{k_r^{(0)}}{r}\right)^2  {\tilde{\Phi}_G}' = 0.
\end{split}
\end{equation} 
The four equations Eqs.(\ref{eq:b1}), (\ref{eq:b2}), (\ref{eq:b6}), and (\ref{eq:b7}) are basic equations 
to be solved for deriving $\tilde{\Phi}_G, \Phi_6, \Phi_h$ and $\Phi_H$.
In this system of equations, two wave-numbers $k_r^{(0)}$ and $k_R^{(0)}$ appear, and 
so the solutions are functions of these two wave-numbers.

\subsection{Equations with respect to $k_r^{(0)} \tau^{4/3}$ and their approximate solutions} 
Furthermore let us define $x$ by $k_r^{(0)} d \tau = r dx$ or 
\begin{equation}
  \label{eq:b10}
 x = \frac{3}{4r_0} k_r^{(0)} \tau^{4/3}
\end{equation} 
for the convenience of calculations. Then the above four equations 
are expressed using $x$ as
\begin{equation}
  \label{eq:b11}
\Phi_{Gx,xx}  =  -\frac{1}{x} \Phi_{Gx,x} - (1 +\frac{\mu}{x} - \frac{1/4}{x^2}) \Phi_{Gx}
+ x^{-1/2}  (2\Phi_h + 2\Phi_H - \frac{2}{3} x^{-3/4} \Phi_{6x}),
\end{equation} 
\begin{equation}
  \label{eq:b12}
\Phi_{6x,x}  =  -\frac{3}{4x} \Phi_{6x} + 3x^{3/4} (\Phi_{h,x} + \frac{2}{x} \Phi_h 
+ \Phi_{H,x} +\frac{1}{x} \Phi_H) 
 +\frac{3}{2} x^{1/4}  (1 - \frac{\mu}{x}) \Phi_{Gx},
\end{equation} 
\begin{equation}
  \label{eq:b13}
      \begin{split}
\Phi_{h,xx}  &=  \frac{1}{3x} \Phi_{h,x} -\frac{1}{3} (\frac{1}{3} +\frac{4\mu}{x} - \frac{20}{x^2}) \Phi_h\\
&+ \frac{10}{3x}\Phi_{H,x} + \frac{1}{3}  (4 - \frac{\mu}{3x} +\frac{10}{x^2}) \Phi_H
- \frac{2}{9} x^{-3/4}  (1 -\frac{\mu}{x} + 6x^{-2}) \Phi_{6x}\\
&- [\frac{4}{9} \frac{\mu}{x^{1/2}} - x^{-3/2}  (\frac{23}{12} 
- \frac{5}{3}\frac{\mu}{x})] \Phi_{Gx} 
 +\frac{1}{2} x^{-3/2}\mu \Phi_{Gx,x},
\end{split}
\end{equation} 
and
\begin{equation}
  \label{eq:b14}
      \begin{split}
\Phi_{H,xx}  &=  -\frac{19}{6x} \Phi_{H,x} -\frac{1}{3} (2 + \frac{\mu}{3x}+
 \frac{7/2}{x^2}) \Phi_H\\
&- \frac{7}{6x}\Phi_{h,x} + \frac{1}{3}  (-\frac{1}{3} + \frac{2\mu}{x} -\frac{7}{x^2}) \Phi_h
+ \frac{1}{9} x^{-3/4}  (1 -\frac{\mu}{x} + \frac{33}{4}x^{-2}) \Phi_{6x}\\
&- [\frac{4}{9} \frac{\mu}{x^{1/2}} + x^{-3/2}  (\frac{5}{6} 
- \frac{7}{12}\frac{\mu}{x})] \Phi_{Gx} 
 -\frac{1}{2} x^{-1/2} \Phi_{Gx,x},
\end{split}
\end{equation} 
where
\begin{equation}
 \label{eq:b15}
\Phi_{Gx} \equiv (\frac{4}{3} k_r^{(0)3} /(r_0)^3)^{1/2} \tilde{\Phi}_G,  
\end{equation} 
\begin{equation}
 \label{eq:b16}
\Phi_{6x} \equiv  (\frac{3}{4} k_r^{(0)} /r_0)^{3/4} \Phi_6,
\end{equation} 
\begin{equation}
 \label{eq:b17}
 \mu \equiv \frac{3}{4} (k_R^{(0)2}/k_r^{(0)})(r_0 /(R_0)^2),  
\end{equation} 
$r = r_0 \tau^{-1/3}, \ R = R_0 \tau^{1/3}$ and $,x$ denotes $d/dx$. 
From Eqs. (\ref{eq:b10}) and (\ref{eq:b17}),
we have
\begin{equation}
 \label{eq:b18}
 \mu/x  = (k_R^{(0)}/k_r^{(0)})^2 (r_0 /R_0)^2 \tau^{-4/3} = [(k_R^{(0)}/R)/(k_r^{(0)}/r)]^2.
\end{equation} 

In the case of small $\tau$ and $x \ll 1$, 
four equations Eqs.(\ref{eq:b11}), (\ref{eq:b12}), (\ref{eq:b13}), and (\ref{eq:b14}) are 
found to have a special set of solutions
\begin{equation}
  \label{eq:b8}
      \begin{split}
\Phi_{6x} &= 0, \quad \Phi_{Gx} = -2 x^{-1/2} x_i^2 \ln \frac{x}{x_i} \ \Phi_{hi} +
2 x^{1/2} x_i \ln \frac{x}{x_i} \ \Phi_{Hi} \\
\Phi_h &= (x/x_i)^{-2}  \Phi_{hi}, \quad \Phi_H =  (x/x_i)^{-1}  \Phi_{Hi},
 \end{split}
\end{equation} 
where $\Phi_{hi}, \Phi_{Hi}, $ and $x_i$ are the values of $\Phi_h, \Phi_H,$ and $x$ at the epoch $\tau = \tau_i$.

The above four equations can be solved from the epoch $x_i$ numerically in the direction of
increasing $x$ (or decreasing $t$), when we use 
Eq. (\ref{eq:b8}) as the condition at the epoch. They were solved using the Runge-Kutta
 method. An example of the numerical solutions is shown in Fig. 1, which gives $\Phi_h$ and
  $\Phi_H$ in the case of $x_i = 0.1$ and $\mu = 0$.
It is found that, during the stage of $x \ll 1$, $\Phi_{h}$ and $\Phi_{H}$ behave simply as
 $\propto x^{-2}$ and 
$\propto x^{-1}$, respectively, but, as $x$ \ (or $\tau$) increases, the behaviors change, and 
they oscillate when $x \gg 1$.

\begin{figure}[t]
\caption{\label{fig:1} Curvature perturbations in the case of $x_i = 0.1$ and $\mu = 0$.
The solid and dotted curves denote $\Phi_h$ and $\Phi_H$, respectively.} 
\centerline{\includegraphics[width=10cm]{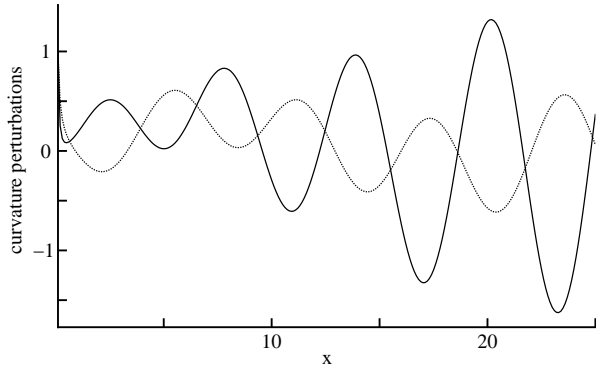}}
\end{figure}

Now let us consider the case when $x \gg 1$ and $\mu/x \ll 1$. In this case, the outer 
wave-number ($k_r^{(0)}/r)$ is $\gg 1$ and much larger than the inner wave-number
($k_R^{(0)}/R$), as can be found from Eq. (\ref{eq:b18}).
 Here we neglect the terms $ \mu/x $  in four equations (\ref{eq:b11}) - (\ref{eq:b14}),
and assume that all quantities have an oscillatory factor $\exp{i\omega x}$, where $\omega$ is 
a constant frequency.  Then $\Phi_{Gx}, \Phi_{6x}, \Phi_h,$ and $\Phi_H$ are expressed as
\begin{equation}
 \label{eq:b19}
       \begin{split}
\Phi_{Gx} &= \Phi_{G0} (x) \exp{i\omega x}, \quad \Phi_{6x} = \Phi_{60} (x) \exp{i\omega x},\\
\Phi_{h} &= \Phi_{h0} (x) \exp{i\omega x}, \quad \Phi_{H} = \Phi_{H0} (x) \exp{i\omega x},
\end{split}
\end{equation} 
where $\Phi_{G0} (x), \Phi_{60} (x), \Phi_{h0} (x),$ and $\Phi_{H0} (x)$ are slowly varying 
monotonic functions. The special case when the quantities cannot be expanded is 
separately treated. 

From Eq.  (\ref{eq:b11}), we obtain for main terms with respect to $1/x \ (\ll 1)$
\begin{equation}
 \label{eq:b20}
(1 - \omega^2 + i\omega x^{-1}) \Phi_{G0} +2i\omega \Phi_{G0,x} = x^{-1/2} 
(2\Phi_{h0} +2\Phi_{H0} -\frac{2}{3} x^{-3/4} \Phi_{60})
\end{equation} 
and similarly from Eq.(\ref{eq:b12})
\begin{equation}
 \label{eq:b21}
\Phi_{h0} +\Phi_{H0} -\frac{1}{3} x^{-3/4} \Phi_{60} = \frac{1}{2} ix^{-1/2} \omega^{-1} 
\Phi_{G0} .
\end{equation} 
From the compatibility of these two equations, we find that%
\begin{equation}
 \label{eq:b22}
\omega = 1 \quad \rm{or} \quad \Phi_{G0} = 0.
\end{equation} 

\subsubsection{The case of $\omega = 1$}
Equations for $\Phi_{h0}, \Phi_{H0}, \Phi_{60},$ and $\Phi_{G0}$ are derived from Eqs. 
 (\ref{eq:b11}) -  (\ref{eq:b14}), solved in Appendix B, and their solutions are expressed as
\begin{equation}
 \label{eq:b23}
       \begin{split}
\Phi_{H0} &= -\frac{1}{3} \Phi_{h0}, \quad  \Phi_{60} = 2 x^{3/4} \Phi_{h0},  \\
\Phi_{G0} &= \frac{8}{3} x^{1/2} (\Phi_{h0})_{,x} , \quad  \Phi_{h0} \propto x^\alpha,
\end{split}
\end{equation} 
where a constant $\alpha$ satisfies 
\begin{equation}
 \label{eq:b24}
\alpha^2 -  \frac{1}{2} \alpha - \frac{1}{4} = 0 
\end{equation} 
and its solutions are 
\begin{equation}
 \label{eq:b25}
\alpha =  \frac{1}{4} (1 \pm \sqrt{5}) \simeq 0.81, -0.31.
\end{equation} 

\subsubsection{The case of $\omega \neq 1$ and $\Phi_{G0} = 0$}
From Eqs.  (\ref{eq:b12}) - (\ref{eq:b14}), we obtain for the lowest-order terms with respect 
to $1/x$
\begin{equation}
 \label{eq:b26}
       \begin{split}
x^{-3/4} \Phi_{60} &= 3(\Phi_{h0} + \Phi_{H0}),  \\
(\frac{7}{9} - \omega^2) \Phi_{h0} &= \frac{2}{3} \Phi_{H0}, \quad
(\frac{1}{3} - \omega^2) \Phi_{H0} = \frac{2}{9} \Phi_{h0},
\end{split}
\end{equation} 
and from the latter two equations, we obtain
\begin{equation}
 \label{eq:b27}
(\omega^2 - 7/9)(\omega^2 -1/3 ) = 4/27,
\end{equation} 
or
\begin{equation}
 \label{eq:b28}
\omega = 1/3,
\end{equation} 
and from one of the above equations we obtain
\begin{equation}
 \label{eq:b29}
 \Phi_{h0} =  \Phi_{H0} .
\end{equation} 
As shown in Appendix B, $\Phi_{h0}$ for $\omega = 1/3$ has the $x$-dependence  as
\begin{equation}
 \label{eq:b30}
 \Phi_{h0} \propto  x^{-7/6}.
\end{equation} 

\subsubsection{The solutions with power dependences}
Let us assume  that
\begin{equation}
 \label{eq:b31}
       \begin{split}
\Phi_{Gx} &= \Phi_{G0} x^{\alpha_G}, \quad \Phi_{6x} = \Phi_{60} x^{\alpha_6},\\
\Phi_{h} &= \Phi_{h0} x^{\alpha_h},  \quad \Phi_{H} = \Phi_{H0} x^{\alpha_H},
\end{split}
\end{equation} 
where $(\alpha_G, \alpha_6, \alpha_h, \alpha_H)$ and $(\Phi_{G0}, \Phi_{60}, \Phi_{h0}, 
\Phi_{H0})$ are constants. Then, as shown in Appendix B, we obtain
\begin{equation}
 \label{eq:b32}
 \Phi_{h0}  = 0, \ \Phi_{60} = 6 \Phi_{H0},\ {\rm{and}} \ \Phi_{G0} = -2 \Phi_{H0},
\end{equation} 
\begin{equation}
 \label{eq:b33}
\alpha_h = \alpha_H = -3, \ \alpha_6 = -9/4, \  {\rm{and}} \ \alpha_G = -7/2.
\end{equation} 
The important character of this type of perturbation is that we have no curvature perturbations 
in the outer space ($\Phi_{h0}  = 0$).

\subsection{Equations with respect to $k_R^{(0)} \tau^{2/3}$ and their approximate solutions}
Moreover let us define $y$ by $k_R^{(0)} d \tau = R dy$ or 
\begin{equation}
  \label{eq:b34}
 y = \frac{3}{2R_0} k_R^{(0)} \tau ^{2/3}
\end{equation} 
similarly to Eq.(\ref{eq:b10}).
Then the above four equations Eqs.(\ref{eq:b1}), (\ref{eq:b2}), (\ref{eq:b6}), and (\ref{eq:b7})
are expressed as
\begin{equation}
  \label{eq:b35}
\Phi_{Gy,yy}  =  -\frac{1}{y} \Phi_{Gy,y} - (1 +\frac{2\nu}{3}y^2 - \frac{1}{y^2}) \Phi_{Gy}
+ y (2\Phi_h + 2\Phi_H - \frac{2}{3} y^{-3/2} \Phi_{6y}),
\end{equation} 
\begin{equation}
  \label{eq:b36}
\Phi_{6y,y}  =  -\frac{3}{2y} \Phi_{6y} + 3y^{3/2} (\Phi_{h,y} + \frac{4}{y} \Phi_h 
+ \Phi_{H,y} +\frac{2}{y} \Phi_H)
 - 3 y^{-1/2} (1 - \frac{2}{3}\nu y^2) \Phi_{Gy},
\end{equation} 
\begin{equation}
  \label{eq:b37}
      \begin{split}
\Phi_{h,yy}  &=  \frac{5}{3y} \Phi_{h,x} -\frac{1}{3}(4 +\frac{2}{9}\nu y^2 -\frac{80}{y^2}) \Phi_h\\
&+ \frac{20}{3y}\Phi_{H,y} - \frac{1}{3} (\frac{1}{3} - \frac{8}{3}\nu y^2 -\frac{40}{y^2}) \Phi_H
+ \frac{2}{9} y^{-3/2} (1 -\frac{2}{3}\nu y^2 - 24 y^{-2}) \Phi_{6y}\\
&-[\frac{8}{27} \nu y + \frac{1}{3} y^{-3} (23 - \frac{40}{3} \nu y^2)] \Phi_{Gy} 
  + y^{-2}  \Phi_{Gy,y},
\end{split}
\end{equation} 
and
\begin{equation}
  \label{eq:b38}
      \begin{split}
\Phi_{H,yy}  &=  -\frac{16}{3y} \Phi_{H,y} -\frac{1}{3}(\frac{1}{3}+ \frac{8}{9}\nu y^2 
 +\frac{14}{y^2}) \Phi_H\\
&- \frac{7}{3y}\Phi_{h,y} + \frac{1}{3} (1 - \frac{4}{9}\nu y^2 - \frac{28}{y^2}) \Phi_h
- \frac{1}{9} y^{-3/2} (1 -\frac{2}{3}\nu y^2 - \frac{33}{y^2}) \Phi_{6y}\\
&-[\frac{8}{27} \nu y -\frac{1}{3} y^{-3} (7- \frac{40}{9} \nu y^2)] \Phi_{Gy} 
 - \frac{2}{3} \nu \Phi_{Gy,y},
\end{split}
\end{equation} 
where
\begin{equation}
 \label{eq:b39}
\Phi_{Gy} \equiv (\frac{4}{3} k_R^{(0)4} /(r_0)^3)^{1/2} \tilde{\Phi}_G,  
\end{equation} 
\begin{equation}
 \label{eq:b40}
\Phi_{6y} \equiv  (\frac{3}{4} k_R^{(0)} /r_0)^{3/4} \tilde{\Phi}_6,
\end{equation} 
\begin{equation}
 \label{eq:b41}
 \nu \equiv  (k_r^{(0)}/k_R^{(0)2})^2 ((R_0)^2 /r_0)^2 = (\frac{3}{4\mu})^2,
\end{equation} 
and $,y$ denotes $d/dy$. From Eqs. (\ref{eq:b34}) and (\ref{eq:b41}), we have
\begin{equation}
 \label{eq:b42}
 \nu y^2  = (k_r^{(0)}/k_R^{(0)})^2 (R_0 /r_0)^2 \tau^{4/3} = [(k_r^{(0)}/r)/(k_R^{(0)}/R)]^2 
 = x/\mu.
\end{equation} 

In the case of small $\tau$ and $y \ll 1$, these four equations  Eqs.(\ref{eq:b35}), (\ref{eq:b36}),
 (\ref{eq:b37}), and (\ref{eq:b38})  have a special solution 
\begin{equation}
  \label{eq:b9}
      \begin{split}
\Phi_{6y} &= 0, \quad \Phi_{Gy} = - y^{-1} y_i^4 \ln \frac{y}{y_i} \ \Phi_{hi} +
 y y_i^2 \ln \frac{y}{y_i} \ \Phi_{Hi} \\
\Phi_h &= (y/y_i)^{-4}  \Phi_{hi}, \quad \Phi_H =  (y/y_i)^{-2}  \Phi_{Hi},
 \end{split}
\end{equation} 
where $\Phi_{hi}, \Phi_{Hi}, $ and $y_i$ are the values of $\Phi_h, \Phi_H,$ and $y$ at the epoch 
$\tau = \tau_i$.
This is found to be identical with the special solution Eq. (\ref{eq:b8}).
The above four equations were solved numerically using Eq. (\ref{eq:b9}) as the condition
at the epoch $y_i$ in the direction of increasing $y$ (or decreasing $t$). 
An example of it is shown in Fig. 2, which gives $\Phi_h$ and $\Phi_H$ 
in the case of $y_i = 0.2$ and $\nu = 0$.

\begin{figure}[t]
\caption{\label{fig:2} Curvature perturbations in the case of $y_i = 0.2$ and $\nu = 0$.
The solid and dotted curves denote $\Phi_h$ and $\Phi_H$, respectively.} 
\centerline{\includegraphics[width=10cm]{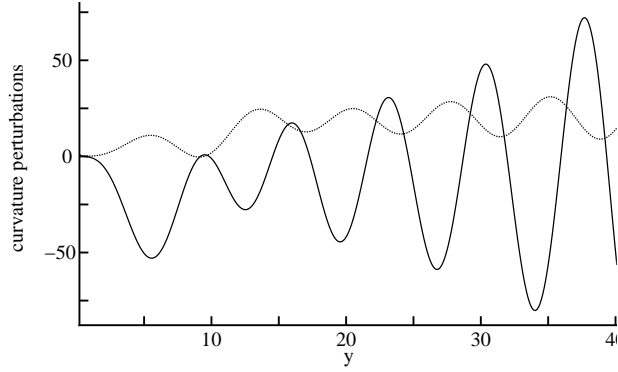}}
\end{figure}

Now let us consider the case when $y \gg 1$ and $\nu y^2 \ (= \mu/x) \ \ll 1$. 
In this case, the inner wave-number ($k_R^{(0)}/R$) is $\gg 1$ and much larger than the outer wave-number
($k_r^{(0)}/r$), as can be found from Eq. (\ref{eq:b42}). Here we neglect the terms $\nu y^2$  
in the four equations,
and assume that all quantities have an oscillatory factor $\exp{i\omega y}$, where $\omega$ is 
a frequency.  Then $\Phi_{Gy}, \Phi_{6y}, \Phi_h,$ and $\Phi_H$ are expressed as
\begin{equation}
 \label{eq:b43}
       \begin{split}
\Phi_{Gy} &= \Phi_{G0} (y) \exp{i\omega y}, \quad \Phi_{6y} = \Phi_{60} (y) \exp{i\omega y},\\
\Phi_{h} &= \Phi_{h0} (y) \exp{i\omega y}, \quad \Phi_{H} = \Phi_{H0} (y) \exp{i\omega y},
\end{split}
\end{equation} 
where $\Phi_{G0} (y), \Phi_{60} (y), \Phi_{h0} (y),$ and $\Phi_{H0} (y)$ are slowly varying 
monotonic functions. The special case when the quantities cannot be expanded is 
treated separately. 

From Eq.  (\ref{eq:b35}), we obtain for the main terms with respect to $1/y \ (\ll 1)$
\begin{equation}
 \label{eq:b44}
(1 - \omega^2 + i\omega y^{-1}) \Phi_{G0} +2i\omega \Phi_{G0,y} = y 
(2\Phi_{h0} +2\Phi_{H0} -\frac{2}{3} y^{-3/2} \Phi_{60})
\end{equation} 
and, similarly, from Eq.(\ref{eq:b36})
\begin{equation}
 \label{eq:b45}
\Phi_{h0} +\Phi_{H0} -\frac{1}{3} y^{-3/2} \Phi_{60} = - iy^{-2} \omega^{-1} \Phi_{G0} .
\end{equation} 
From the compatibility of Eqs. (\ref{eq:b44})  and  (\ref{eq:b45}), we find that
\begin{equation}
 \label{eq:b46}
\omega = 1 \quad \rm{or} \quad \Phi_{G0} = 0.
\end{equation} 

\subsubsection{The case of $\omega = 1$}
The equations for $\Phi_{h0}, \Phi_{H0}, \Phi_{60},$ and $\Phi_{G0}$ are derived from Eqs. 
 (\ref{eq:b35}) -  (\ref{eq:b38}), solved in Appendix C, and their solutions are expressed as
\begin{equation}
 \label{eq:b47}
       \begin{split}
\Phi_{H0} &= -\frac{1}{10} \Phi_{h0}, \quad  \Phi_{60} = \frac{11}{5}  y^{3/2} \Phi_{h0},  \\
\Phi_{G0} &= \frac{1}{6} i y^2 \Phi_{h0} , \quad  \Phi_{h0} \propto y^{-7/2}.
\end{split}
\end{equation} 

\subsubsection{The case of $\omega \neq 1$ and $\Phi_{G0} = 0$}
From Eqs.  (\ref{eq:b36}) - (\ref{eq:b38}), we obtain for the lowest-order terms with respect 
to $1/y$
\begin{equation}
 \label{eq:b48}
       \begin{split}
y^{-3/2} \Phi_{60} &= 3(\Phi_{h0} + \Phi_{H0}),  \\
(\frac{2}{3} - \omega^2) \Phi_{h0} &= \frac{5}{9} \Phi_{H0}, \\
(\frac{4}{9} - \omega^2) \Phi_{H0} &= 0.
\end{split}
\end{equation} 
From the latter two equations, we find that there are the following two cases (a) and (b):

\noindent (a) $\omega = 2/3$ and $ \Phi_{H0} = \frac{2}{5} \Phi_{h0} $

As shown in Appendix C, $\Phi_{h0}$ for $\omega = 2/3$ has  $y$-dependence  as
\begin{equation}
 \label{eq:b51}
 \Phi_{h0} \propto  y^\alpha,
\end{equation} 
where a constant $\alpha$ satisfies 
\begin{equation}
 \label{eq:b52}
 \alpha^2 + \frac{653}{96} \alpha + \frac{1097}{64} = 0,
\end{equation} 
and its solution is 
\begin{equation}
 \label{eq:b53}
 \alpha = \alpha_r \pm i \alpha_m 
\end{equation} 
with $\alpha_r \simeq -4.40$ and $\alpha_m \simeq 2.36$. Then $\Phi_{h0}$ can also be 
expressed as $ \Phi_{h0} \propto y^{\alpha_r} \exp ({\pm i\alpha_m \ln y})$, which 
shows that $\Phi_{h0}$  oscillates slowly in a logarithmic way.

\noindent (b) $\omega = (2/3)^{1/2}$ and $\Phi_{H0} = 0$

As shown in Appendix C, $\Phi_{h0}$ for $\omega = (2/3)^{1/2}$ has the $y$-dependence  as
\begin{equation}
 \label{eq:b51a}
 \Phi_{h0} \propto  y^\beta,
\end{equation} 
where a constant $\beta$ satisfies 
\begin{equation}
 \label{eq:cc16}
 \beta^2 + \frac{33}{16} \beta - \frac{285}{32} = 0,
\end{equation} 
and its solution is 
\begin{equation}
 \label{eq:cc17}
 \beta \simeq 2.13, \ -4.19.
\end{equation} 

\subsubsection{The solutions with power dependences}
Let us assume  that
\begin{equation}
 \label{eq:b54}
       \begin{split}
\Phi_{Gy} &= \Phi_{G0} y^{\alpha_G}, \quad \Phi_{6y} = \Phi_{60} y^{\alpha_6},\\
\Phi_{h} &= \Phi_{h0} y^{\alpha_h},  \quad \Phi_{H} = \Phi_{H0} y^{\alpha_H},
\end{split}
\end{equation} 
where $(\alpha_G, \alpha_6, \alpha_h, \alpha_H)$ and $(\Phi_{G0}, \Phi_{60}, \Phi_{h0}, 
\Phi_{H0})$ are constants. Then, as shown in Appendix C, we obtain
\begin{equation}
 \label{eq:b55}
 \Phi_{H0}  = -2 \Phi_{h0}, \quad \Phi_{60} = 5 \Phi_{h0},\quad {\rm{and}} \quad \Phi_{G0} = 
 -\frac{16}{3} \Phi_{h0},
\end{equation} 
\begin{equation}
 \label{eq:b56}
\alpha_h = \alpha_H = 1/8, \quad \alpha_6 = 13/8, \quad  {\rm{and}} \quad \alpha_G = 9/8.
\end{equation} 
In this case, we have both non-zero perturbations in the inner and outer spaces, in different from 
Eq.(\ref{eq:b32}).

\subsection{Summary of approximate solutions in the scalar mode}
In the case of small $\tau$ and $x \ll 1$ \ (or $y \ll 1$), we have $\Phi_h \propto x^{-2}$ and 
$\Phi_H \propto x^{-1}$ \ (or $\Phi_h \propto y^{-4}$ and $\Phi_H \propto y^{-2}$), respectively, 
where  $x \equiv \frac{3}{4r_0} k_r^{(0)} \tau^{4/3}$ and $y \equiv \frac{3}{2R_0} k_R^{(0)} \tau^{2/3}$.

In the case of $x \gg 1$ and $\mu/x \ll 1$ \ (or $y \gg 1$ and  $\nu y^2 \ (= x/\mu) \ \ll 1$), 
$k_r$ dependent perturbations (or $k_R$ dependent perturbations) dominate $k_R$ dependent 
ones (or $k_r$ dependent ones), respectively. Among the solutions in these cases, 
only the solutions in the cases of $\omega = 1, \ \alpha \simeq 0.81$ \ (of $x \gg 1, \ 
\mu/x \ll 1$) and $\omega = (2/3)^{1/2}, \ \beta \simeq 2.13$ \ (of $y \gg 1, \ 
\nu y^2 \ll 1$) grow in the direction of increasing $\tau$, and all the other 
perturbations grow in the direction of decreasing $\tau$ (or increasing $t$).

In  Figs.1 and 2, we can see the behaviors of the solutions with $\omega = 1, \ \alpha \simeq
 0.81$ in the region of $x \gg 1$ and $\omega = (2/3)^{1/2}, \ \beta \simeq 2.13$ \ 
 in the region of $y \gg 1$. It is noted that only these perturbations  appear as 
 growing ones, when we solve in the direction of increasing $x$ or $y$. 

\section{Solutions in the vector mode}
We consider here the VS, SV and VV cases at the final stage  with $r = r_0 \tau^{-1/3}$ and 
$R = R_0 \tau^{1/3}$, and $\rho = p = 0$.

\subsection{The VS case}
From Eqs. (\ref{eq:a47}) and (\ref{eq:a48}) , we obtain
\begin{equation}
 \label{eq:c1}
{\psi_r}'' + \frac{7/3}{\tau} {\psi_r}' + [(\frac{k_r^{(1)}}{r})^2 + (\frac{k_R^{(0)}}{R})^2] \psi_r = 0, 
\end{equation} 
\begin{equation}
 \label{eq:c2}
J_r = \frac{R}{k_R^{(0)}} ({\Psi_r}' + \frac{4/3}{\tau} \Psi_r), 
\end{equation}
where $' = \partial/\partial \tau$ and Eq. (\ref{eq:a49}) is derived from Eqs. (\ref{eq:a47}) 
and (\ref{eq:a48}).

For $(\frac{k_r^{(1)}}{r})^2 + (\frac{k_R^{(0)}}{R})^2 \ll \tau^{-2}$, we have two solutions
\begin{equation}
 \label{eq:c3}
\Psi_r \propto 1, \ \tau^{-4/3}, \quad J_r = \tau^{-2/3}, \ 0, 
\end{equation}
respectively. On the other hand, for $(\frac{k_r^{(1)}}{r})^2 + (\frac{k_R^{(0)}}{R})^2 \gg
 \tau^{-2}$, we can consider two different cases $(\frac{k_r^{(1)}}{r})^2 \gg (\frac{k_R^{(0)}}{R})^2$
 and $(\frac{k_r^{(1)}}{r})^2 \ll (\frac{k_R^{(0)}}{R})^2$.
 
 If we use $x$ defined by $k_r^{(1)} d\tau = r dx$ or $x = (3/4r_0) k_r^{(1)} \tau^{4/3}$, we obtain
 from Eq.(\ref{eq:c1})
\begin{equation}
 \label{eq:c4}
\Psi_{r,xx} + \frac{2}{x} \Psi_{r,x} + (1 + \mu/x) \Psi_r = 0, 
\end{equation} 
where $\mu/x = (k_R^{(0)}/k_r^{(1)})^2(r_0/R_0)^2 \tau^{-4/3} = [(k_R^{(0)}/R)/(k_r^{(1)}/r)]^2$.

For $\Psi_r = x^{-1/2} \tilde{\Psi}_r$, this equation is replaced by
\begin{equation}
 \label{eq:c5}
\tilde{\Psi}_{r,xx} + \frac{1}{x} \tilde{\Psi}_{r,x} + [-\frac{1}{4} x^{-2} +1 + \mu/x] \tilde{\Psi}_r = 0, 
\end{equation} 
so that 
\begin{equation}
 \label{eq:c6}
\tilde{\Psi}_r \simeq J_0 (x) \ {\rm or} \ N_0(x) 
\end{equation} 
for $x \gg 1$ and $\mu/x \ll 1$, where $J_0(x)$ and $N_0(x)$ are the $0$ th 
Bessel functions of first and second kinds, respectively. Here $J_0(x)$ and $N_0(x)$ for 
$x \gg 1$ have asymptotic behaviors such as 
\begin{equation}
 \label{eq:c6a}
J_0 \simeq \sqrt{\frac{2}{\pi x}} \cos(x - \pi/4), \quad  N_0 \simeq \sqrt{\frac{2}{\pi x}} 
\sin(x - \pi/4).
\end{equation} 

If we use $y$ defined by Eq. (\ref{eq:b34}), we obtain 
\begin{equation}
 \label{eq:c7}
\Psi_{r,yy} + \frac{3}{y} \Psi_{r,y} + (1 + \nu y^2) \Psi_r = 0, 
\end{equation} 
where $\nu y^2 = (k_r^{(1)}/k_R^{(0)})^2(R_0/r_0)^2 \tau^{4/3} = [(k_r^{(1)}/r)/(k_R^{(0)}/R)]^2
\ (= x/\mu)$.
For $\Psi_r = y^{-1} \tilde{\Psi}_r$, this equation is
\begin{equation}
 \label{eq:c8}
\tilde{\Psi}_{r,yy} + \frac{1}{y} \tilde{\Psi}_{r,y} + [- y^{-2} +1 + \nu y^2] \tilde{\Psi}_r = 0, 
\end{equation} 
so that 
\begin{equation}
 \label{eq:c9}
\tilde{\Psi}_r \simeq J_0 (y) \ {\rm or} \ N_0(y) 
\end{equation} 
for $ y \gg 1$ and $\nu y^2 \ll 1$. 

It is found, therefore,  that for 
$(k_r^{(1)}/r)^2 + (k_R^{(0)}/R)^2 \gg \tau^{-2}$, $\psi_r$ takes the wavy behaviors 
with respect to $x$ or $y$. From the asymptotic behaviors of Bessel functions,
 the amplitudes of $\psi_r$ are found to be $\sqrt{\frac{2}{\pi x}} x^{-1/2}$ and 
 $\sqrt{\frac{2}{\pi y}} y^{-1}$ for $\mu/x \ll 1$ and $\nu y^2 (= x/\mu) \ll 1$, respectively.
 
 In terms of $x$ and $y$, solutions (\ref{eq:c3}) are expressed as $\Psi_r \propto 1, x^{-1}
 \ ($ for $x \ll 1)$ \ or \ $\propto 1, y^{-2} \ $(for $y \ll 1)$.

\subsection{The SV case}
From Eqs. (\ref{eq:a50}) and (\ref{eq:a51}) , we obtain
\begin{equation}
 \label{eq:c10}
{\Psi_R}'' - \frac{1/3}{\tau} {\Psi_R}' + [(\frac{k_r^{(0)}}{r})^2 + (\frac{k_R^{(1)}}{R})^2 
- \frac{4/3}{\tau^2}] \Psi_R = 0, 
\end{equation} 
\begin{equation}
 \label{eq:c11}
J_R = \frac{r}{2 k_r^{(0)}} ({\Psi_R}' + \frac{2/3}{\tau} \Psi_R), 
\end{equation}
where Eq. (\ref{eq:a52}) is derived from Eqs. (\ref{eq:a50}) and (\ref{eq:a51}).

For $(\frac{k_r^{(0)}}{r})^2 + (\frac{k_R^{(1)}}{R})^2 \ll \tau^{-2}$, we have two solutions
\begin{equation}
 \label{eq:c12}
\Psi_R \propto \tau^2, \ \tau^{-2/3}, \quad J_R = \tau^{2/3}, \ 0, 
\end{equation}
respectively. On the other hand, for $(\frac{k_r^{(0)}}{r})^2 + (\frac{k_R^{(1)}}{R})^2 \gg
 \tau^{-2}$, we can consider two different cases $(\frac{k_r^{(0)}}{r})^2 \gg (\frac{k_R^{(1)}}{R})^2$
 and $(\frac{k_r^{(0)}}{r})^2 \ll (\frac{k_R^{(1)}}{R})^2$.
 
 By use of $x$ defined by $k_r^{(0)} d\tau = r dx$ or $x = (3/4r_0) k_r^{(0)} \tau^{4/3}$, we obtain
\begin{equation}
 \label{eq:c13}
\Psi_{R,xx}  + (1 + \mu/x -\frac{3/4}{x^2}) \Psi_R = 0, 
\end{equation} 
where $\mu/x = (k_R^{(1)}/k_r^{(0)})^2 (r_0/R_0)^2 \tau^{-4/3} = [(k_R^{(1)}/R)/(k_r^{(0)}/r)]^2$.

For $\Psi_R = x^{1/2} \tilde{\Psi}_R$, this equation is replaced by
\begin{equation}
 \label{eq:c14}
\tilde{\Psi}_{R,xx} + \frac{1}{x} \tilde{\Psi}_{R,x} + [- x^{-2} +1 + \mu/x] \tilde{\Psi}_R = 0, 
\end{equation} 
so that 
\begin{equation}
 \label{eq:c15}
\tilde{\Psi}_R \simeq J_0 (x) \ {\rm or} \ N_0(x) 
\end{equation} 
for $ x \gg 1$ and $\mu/x \ll 1$.

If we use $y$ defined by $k_R^{(1)} d\tau = R dy$ or $y = (3/2R_0) k_R^{(1)} \tau^{2/3}$,
 we obtain 
\begin{equation}
 \label{eq:c16}
\Psi_{R,yy} - \frac{1}{y} \Psi_{R,y} + (-\frac{3}{y^2} +1 + \nu y^2) \Psi_R = 0, 
\end{equation} 
where $\nu y^2 = (k_r^{(0)}/k_R^{(1)})^2(R_0/r_0)^2 \tau^{4/3} = [(k_r^{(0)}/r)/(k_R^{(1)}/R)]^2$.
For $\Psi_r = y \tilde{\Psi}_r$, this equation is
\begin{equation}
 \label{eq:c17}
\tilde{\Psi}_{R,yy} + \frac{1}{y} \tilde{\Psi}_{R,y} + [- 4y^{-2} +1 + \nu y^2] \tilde{\Psi}_R = 0, 
\end{equation} 
so that 
\begin{equation}
 \label{eq:c18}
\tilde{\Psi}_R \simeq J_0 (y) \ {\rm or} \ N_0(y) 
\end{equation} 
for $ y \gg 1$ and $\nu y^2 \ll 1$.

From the asymptotic behaviors of Bessel functions,
 the amplitudes of $\psi_R$ are found to be $\sqrt{\frac{2}{\pi x}} x^{1/2}$ and 
 $\sqrt{\frac{2}{\pi y}} y$ for $\mu/x \ll 1$ and $\nu y^2 (= x/\mu) \ll 1$, respectively.
 
  In terms of $x$ and $y$, solutions (\ref{eq:c12}) are expressed as $\Psi_R \propto x^{3/2},
   x^{-1/2} \ ($ for $x \ll 1)$ \ or \ $\propto y^3, y^{-1} \ ($ for $y \ll 1)$.

\subsection{The VV case}
 Equation (\ref{eq:a53}) reduces to 
\begin{equation}
 \label{eq:c19}
{G^{(11)}}'' + \frac{1}{\tau} {G^{(11)}}' + [(\frac{k_r^{(1)}}{r})^2 + (\frac{k_R^{(1)}}{R})^2
- \frac{4/9}{\tau^2}] G^{(11)} = 0.
\end{equation} 
For $\tau \simeq 0$, we have 
\begin{equation}
 \label{eq:c20}
G^{(11)} \propto \tau^{2/3}, \quad \tau^{-2/3}.
\end{equation} 
If we use $x$ or $y$ defined by $x = (3/4r_0) k_r^{(1)} \tau^{4/3}$ or 
$y = (3/2R_0) k_R^{(1)} \tau^{2/3}$, we obtain  
\begin{equation}
 \label{eq:c21}
{G^{(11)}}_{,xx} + \frac{1}{x} {G^{(11)}}_{,x} + [1 + \mu/x + \frac{1}{4x^2}] G^{(11)} = 0,
\end{equation} 
or 
\begin{equation}
 \label{eq:c22}
{G^{(11)}}_{,yy} + \frac{1}{y} {G^{(11)}}_{,y} + [1 + \nu y^2 - \frac{1}{y^2}] G^{(11)} = 0,
\end{equation} 
where $\mu /x$ and $\nu y^2$ are defined by 
$\mu/x = (k_R^{(1)}/k_r^{(1)})^2 (r_0/R_0)^2 \tau^{-4/3} = [(k_R^{(1)}/R)/(k_r^{(1)}/r)]^2$,
 and $\nu y^2  = 1/(\mu/x).$  
 From these equations, therefore, we obtain 
\begin{equation}
 \label{eq:c23}
G^{(11)} = J_0(x) \quad {\rm or}  \quad J_0 (y)
\end{equation} 
for the situation of [$\frac{1}{4x^2} \ll 1$ and $\mu/x \ll 1$] or [$\frac{1}{y^2} \ll 1 $ and
$\nu y^2 \ll 1$], respectively.

From the asymptotic behaviors of Bessel functions,
 the amplitudes of $G^{(11)}$ are found to be $\sqrt{\frac{2}{\pi x}} $ and 
 $\sqrt{\frac{2}{\pi y}} $ for $\mu/x \ll 1$ and $\nu y^2 (= x/\mu) \ll 1$, respectively.
 
   In terms of $x$ and $y$, solutions (\ref{eq:c20}) are expressed as $G^{(11)} \propto x^{1/2},
   x^{-1/2} \ ($ for $x \ll 1)$ \ or \ $\propto y, y^{-1} \ ($ for $y \ll 1)$.

\section{Solutions in the tensor mode}
Here we consider the TS and ST cases at the final stage. 

\subsection {The TS case}
The equation for $h_T^{(2)}$ reduces to
\begin{equation}
 \label{eq:d1}
{h_T^{(2)}}'' + \frac{1}{\tau} {h_T^{(2)}}' + [(\frac{k_r^{(2)}}{r})^2 + (\frac{k_R^{(0)}}{R})^2] 
h_T^{(2)} = 0.
\end{equation} 
Here we use $x$, defined by $k_r^{(2)} d\tau = r dx$ (or $x = (3/4r_0) k_r^{(2)} \tau^{4/3}$),
and $y$, defined by $k_R^{(0)} d\tau = R dy$ (or $y = (3/2R_0) k_R^{(0)} \tau^{2/3}$).
Then we obtain
\begin{equation}
 \label{eq:d2}
{h_T^{(2)}}_{,xx} + \frac{1}{x} {h_T^{(2)}}_{,x} + (1 + \mu/x) h_T^{(2)} = 0.
\end{equation} 
and 
\begin{equation}
 \label{eq:d3}
{h_T^{(2)}}_{,yy} + \frac{1}{y} {h_T^{(2)}}_{,y} + (1 + \nu y^2) h_T^{(2)} = 0,
\end{equation} 
where $\mu/x = (k_R^{(0)}/k_r^{(2)})^2 (r_0/R_0)^2 \tau^{-4/3} = [(k_R^{(0)}/R)/(k_r^{(2)}/r)]^2$,
 and $\nu y^2  = (\mu/x)^{-1}.$  

For $\mu/x \ll 1$ or $\nu y^2 \ll 1$, we have
\begin{equation}
 \label{eq:d4}
h_T^{(2)} = J_0 (x) \quad {\rm or} \quad  J_0 (y),
\end{equation} 
respectively, as the solutions which are regular at the epoch $\tau = 0$, 
corresponding to $x = 0$ or $y = 0$.

From the asymptotic behaviors of Bessel functions,
 the amplitudes of $h_T^{(2)}$ for $x \gg 1$ (or $y \gg 1$) are found to be
 $\sqrt{\frac{2}{\pi x}}$\  for $\mu/x \ll 1$ \ (or\ $\sqrt{\frac{2}{\pi y}} $\ for
  $\nu y^2 (= x/\mu) \ll 1$), respectively.
 
\subsection {The ST case}
Equation for $H_T^{(2)}$ reduces to
\begin{equation}
 \label{eq:d5}
{H_T^{(2)}}'' + \frac{1}{\tau} {H_T^{(2)}}' + [(\frac{k_r^{(0)}}{r})^2 + (\frac{k_R^{(2)}}{R})^2] 
H_T^{(2)} = 0.
\end{equation} 
Here we use $x$  and $y$ which are defined by  $k_r^{(0)} d\tau = r dx$ (or $x = (3/4r_0) 
k_r^{(0)} \tau^{4/3}$) and $y$ defined by $k_R^{(2)} d\tau = R dy$ (or $y = (3/2R_0) k_R^{(2)} \tau^{2/3}$). 
Then we obtain
\begin{equation}
 \label{eq:d6}
{H_T^{(2)}}_{,xx} + \frac{1}{x} {H_T^{(2)}}_{,x} + (1 + \mu/x) H_T^{(2)} = 0
\end{equation} 
and 
\begin{equation}
 \label{eq:d7}
{H_T^{(2)}}_{,yy} + \frac{1}{y} {H_T^{(2)}}_{,y} + (1 + \nu y^2) H_T^{(2)} = 0,
\end{equation} 
where $\mu/x$ and  $\nu y^2  \ (= 1/(\mu/x))$ \ are also given in the TS case. 
For $\mu/x \ll 1$ or $\nu y^2 \ll 1$, we have
\begin{equation}
 \label{eq:d8}
H_T^{(2)} = J_0 (x) \quad {\rm or} \quad  J_0 (y),
\end{equation} 
respectively, as the solutions that are regular at the epoch $\tau = 0$, 
corresponding to $x = 0$ or $y = 0$.

From the asymptotic behaviors of Bessel functions,
 the amplitudes of $H_T^{(2)}$ for $x \gg 1$ (or $y \gg 1$) are found to be 
 $\sqrt{\frac{2}{\pi x}}$  for $\mu/x \ll 1$\ (or \ $\sqrt{\frac{2}{\pi y}} $ \ for 
 $\nu y^2 (= x/\mu) \ll 1$), respectively.

\section{Evolution of models, perturbations and their spectrum}
We consider the evolution of the inner and outer spaces and the perturbations in their 
spaces at the final stage. 
Characteristic epochs in the evolution are as follows.
Since the total universe starts from the isotropic state, we assume first
for simplicity that, at the initial epoch $t_1$ of this final stage, both scale-factors are equal, i.e., 
$R(\tau_1) = r(\tau_1)$, where 
$\tau_1 = t_0 - t_1$. As $R(\tau_1) = R_0\ \cdot  (\tau_1)^{1/3}$ and $r(\tau_1) = r_0\ \cdot 
(\tau_1)^{-1/3}$, we have 
\begin{equation}
 \label{eq:e1}
\tau_1 = (r_0/R_0)^{3/2}.
\end{equation} 

As for the $k_r$ and $k_R$ dependences of perturbations, moreover, we assume 
that initially both spaces have the same spectra (or $k_r$ and $k_R$ dependence)
  with equal mean values $\bar{k}_R = 
\bar{k}_r$.  Then we have the relation $(\bar{k}_R/R)/(\bar{k}_r/r) = 1$ at epoch $\tau_1$.

Secondly, let us consider a characteristic epoch $\tau_2$, when the mean wavelength of 
$k_r$ dependent perturbations in the outer space is equal to the sound wavelength, that is, 
$\bar{k}_r /r(\tau_2) = 1/(c_s \tau_2)$ in all modes, where $c_s$ is the sound velocity of photon
fluids ($\approx 1$). The mean value of $x \ (\equiv \frac{3}{4r_0} k_r \tau^{4/3})$ is 
$\approx 1$ and $\gg 1$ at $\tau = \tau_2$ and $\tau \gg \tau_2$, respectively.

Thirdly, we consider another characteristic epoch $\tau_3$, when the mean wavelength of
$k_R$ dependent perturbations in the outer space is equal to the sound wavelength, i.e., 
$\bar{k}_R/R = 1/(c_s \tau_3)$ in all modes.  The mean values of $y \ (\equiv \frac{3}{2R_0} k_R
 \tau^{2/3})$ are $\approx 1$ and $\gg 1$ at $\tau = \tau_3$ and $\tau \gg \tau_3$, 
 respectively. Then 
\begin{equation}
 \label{eq:e2}
\tau_2 = (r_0/c_s \bar{k}_r)^{3/4} \ {\rm and} \  \tau_3 = (R_0/c_s \bar{k}_R)^{3/2},
\end{equation} 
so that, using Eq. (\ref{eq:e1}), we obtain the relation
\begin{equation}
 \label{eq:e3}
\tau_1 \tau_3 = (\tau_2)^2
\end{equation} 
between the three epochs $\tau_1, \tau_2$ and $\tau_3$.

Finally, we consider as the limiting epoch the epoch $\tau_\dagger$ when the cosmological entropy reaches the 
Guth level and the inner space decouples from the outer space. According to my previous 
work\citep{tom}, we have $R(\tau_\dagger) /R(\tau_M) \simeq 3.0 \times 10^{-13}$, 
where $\tau_M$ represents an epoch at the stage when the inner space has the maximum
expansion. So, if we assume $\tau_M \approx \tau_1$ roughly, 
\begin{equation}
 \label{eq:e3a}
\tau_\dagger / \tau_1 \approx (3.0 \times 10^{-13})^3 \simeq  3.0 \times 10^{-38},
\end{equation} 
because of $R \propto \tau^{1/3}$.

Now we define the spectrum of perturbations in the outer space
as the square of absolute values of curvature perturbations in the scalar mode
(i.e., $|\Phi_h|^2 $)  which is expressed as their functions of $k_r$.  The form of the primeval
spectrum appearing before the final stage is assumed to be a simple one with a peak in the 
mean value ($\bar{k}_r$) of $k_r$.  At the final stage, the spectrum in the case
 of $x\gg 1$ and $\mu/x \ll 1$ changes with time due to the evolution of
 the $k_r$ dependent perturbations, but the spectrum  in the case of
$y \gg 1$ and  $\nu y^2 \ (= x/\mu) \ \ll 1$ does not change with time,
because $k_R$ dependent perturbations are dominant and do not depend on $k_r$.

As we have seen, perturbations are functions of two wave-numbers $k_r$ and $k_R$.
Here let us consider the behaviors of perturbations with various values of $k_r$.
To simplify the treatment about them,
 we fix in the following the value of $k_R$ to the mean value $\bar{k}_R$, 
 when we compare the $k_r$ dependence with the $k_R$ dependence.

\noindent (a)  First we consider the perturbations with the mean wavelengths, i.e., $k_r =  \bar{k}_r$ 
 at epoch $\tau_1$. Then, at epochs $\tau_1 \gg \tau \gg
 \tau_3$, we have $y \gg 1$  and
\begin{equation}
 \label{eq:e4}
(\bar{k}_r/r)/(\bar{k}_R/R) \ll 1\quad  {\rm or} \quad\nu y^2 \ (= x/\mu) \ \ll 1,
\end{equation} 
because $r$ increases and $R$ decreases with time. In this case, from the results in Sect. 3,
it is found that, in the outer space, 
 the $k_R$ dependent perturbations dominate and the change in  $k_r$ dependent 
perturbations is relatively small, because the $k_R$ dependent perturbations have no role to
changing the $k_r$ dependence of the spectrum. 
At epochs $\tau_\dagger < \tau < \tau_3$, we have $y < 1$ 
and the wave-like behaviors disappear in both of $k_r$ and $k_R$ dependent perturbations,
and so the change in perturbations is expressed in the form of power functions 
 of $\tau$.  Accordingly the evolution of perturbations gives only small change in the $k_r$ 
 dependence of the spectrum around $k_r = \bar{k}_r$.

\noindent (b) Next we consider the perturbations with $k_r < \bar{k}_r$.  At epochs $\tau_1 
\gg \tau \gg \tau_3$, we have $y \gg 1$ and
\begin{equation}
 \label{eq:e5}
 (k_r/r)/(\bar{k}_R/R) \ll 1 \quad {\rm or} \quad \nu y^2\ (= x/\mu)   \ll 1
\end{equation} 
similarly for the mean perturbations in the inner space ($k_R =  \bar{k}_R$). So, the 
$k_R$ dependent perturbations dominate and the change in $k_r$ dependent perturbations
 are relatively small. That is, we have small changes in the spectrum for $k_r < \bar{k}_r$.

\noindent (c) Thirdly we consider the perturbations with $k_r  >  \bar{k}_r$. Then we have
\begin{equation}
 \label{eq:e6}
(\bar{k}_R/R)/ (k_r/r) < 1 \quad {\rm or} \quad \mu/x  < 1
\end{equation} 
during the time interval of
\begin{equation}
 \label{eq:e7}
\tau_1 \quad > \quad \tau \quad > \quad \tau_k,
\end{equation} 
where 
\begin{equation}
 \label{eq:e8}
 k_r/r(\tau_k)/[\bar{k}_R/R(\tau_k)] = 1 
\end{equation} 
or 
\begin{equation}
 \label{eq:e9}
\tau_k = (\bar{k}_r/k_r)^{3/2} \tau_1 .
\end{equation} 
In the last equation, we used the relation $\bar{k}_R = \bar{k}_r$.
If $\tau_1 > \tau_k > \tau_2$, we have $x > 1$ and the $k_r$ dependent wavy behavior
in these perturbations dominates the 
$k_R$ dependent perturbations in the outer space for $\tau_1 > \tau > \tau_k$.
If $\tau_2 > \tau_k > \tau_3$, the $k_r$ dependent wavy behavior dominates the $k_R$ 
dependent perturbations for $\tau_1 > \tau > \tau_2$
and the $k_r$ dependent power-like behavior dominates them for $\tau_2 > \tau > \tau_k$.
If $ \tau_3 >\tau_k >\tau_\dagger $,  the $k_r$ dependent perturbations 
dominate the $k_R$ dependent perturbations for $\tau_1 > \tau > \tau_3$, and both of $k_r$ 
and $k_R$ dependent perturbations are expressed for $ \tau_3 > \tau > \tau_k$ 
as power functions of $\tau$ being independent of $k_r$, so that there is no change in 
the spectrum. 

From the above analyses, it is found to be for
\begin{equation}
 \label{eq:e10}
k_r > \bar{k}_r  \quad {\rm and} \quad \tau_k < \tau_3.  
\end{equation} 
that, in the most effective way, the $k_r$ dependent perturbations 
dominate the $k_R$ dependent perturbations and can modify the spectrum of 
perturbations in the outer space. From Eqs. (\ref{eq:e9}) and  (\ref{eq:e10}), 
we obtain the condition
\begin{equation}
 \label{eq:e11}
k_r > k_{rm} \ [\equiv \bar{k}_r  (\tau_1/\tau_3)^{2/3}] ,
\end{equation} 
where $k_{rm}$ is the critical values of $k_r$ with maximum spectral changes and
\begin{equation}
 \label{eq:e12}
1 < (\tau_1/\tau_3)^{2/3} < (\tau_1/\tau_\dagger)^{2/3}  \approx 10^{25}.
\end{equation} 
This condition (\ref{eq:e11}) means that, for larger values of  
$\tau_1/\tau_3 \ (\gg 1)$,  we have spectral changes  for larger $k (> k_{rm})$.

\section{Concluding remarks}
In this paper we studied the evolution of various perturbations (including the curvature
 perturbations) in three modes and found that both of $k_r$ and $k_R$ dependences appear
 in the perturbations. Sometimes the perturbations depend mainly on $k_r$, and sometimes they
 depend mainly on $k_R$. The $k_R$ dependent perturbations do not depend on $k_r$, and 
 so have no influence on the spectrum of $k_r$ dependent perturbations.
 It was found that, in the large interval of $k_r$ including the mean
  value $\bar{k}_r$, the $k_R$ dependent perturbations dominate the $k_r$
 dependent perturbations in the outer space, and so the spectrum of perturbations as
 functions of $k_r$ does not change with time, and that it is $k_r > k_{rm} \ [\equiv \bar{k}_r 
  (\tau_1/\tau_3)^{2/3}] $ that $k_r$ dependent perturbations modify the spectrum 
 by dominating the $k_R$ dependent perturbations, where $\tau_1$ and $\tau_3$ are
 the initial epoch at the final stage and the epoch when the mean wavelength of $k_R$ 
 dependent perturbations is equal to the sound wavelength, respectively, and 
 $\tau_1/\tau_3 \gg 1$. 
 
 Thus, in our classical treatments of perturbations, 
 the spectrum of perturbations in the outer space starts from the primeval one
 at the nearly isotropic stage ($\tau > \tau_1$), changes mainly for $k_r > k_{rm}$ 
 due to the evolution of  perturbations at the final stage ($\tau_1 > \tau > \tau_\dagger$), 
 and transfers to the observed spectrum of perturbations in the Friedmann stage (after epoch 
 $\tau_\dagger$). During the long period from the primeval stage to the final anisotropic stage 
 of the multi-dimensional  universe, the perturbations may be created by the quantum fluctuations 
 of scalar, vector and tensor fields. Their treatments may be very interesting, but are beyond the 
 scope of the present work. 
  
  We have paid attention only to the spectrum in the scalar mode in Sect. 6. 
  Similar discussions on the spectrum in the vector and tensor modes may be possible.
  
  In this paper we have treated the case with no viscous processes. In the cases with 
  viscous processes, the dynamical evolution of perturbations may be much more complicated,
  but the fundamental structure, that, in the outer space,  perturbations depend on both
   $k_r$ and $k_R$ and the change in the spectrum depends on how the $k_r$ dependence 
  dominates the $k_R$ dependence, may be invariant.

\section*{Acknowledgements}
The numerical calculations in this work were carried out on SR16000 at
YITP in Kyoto University.

\appendix
\section{Harmonics and gauge transformations}
\subsection{Harmonics}
The scalar harmonics in the outer and inner spaces are derived from scalar Helmholtz 
equations
\begin{equation}
 \label{eq:aa1}
{q^{(0)}}^{|i}_{|i} + k_r^{(0)2} q^{(0)} = 0, \quad
{Q^{(0)}}^{|a}_{|a} + k_R^{(0)2} Q^{(0)} = 0
\end{equation} 
where $q^{(0)} = q^{(0)} (x^i), \  Q^{(0)} = Q^{(0)} (x^a), \ i=1, ....,d, \ a=d+1, ..., d+D.$
$|i$ and $|a$ denote covariant derivatives with respect to $x^i$ and $X^a$ in the outer and inner 
spaces, respectively.  The vector and tensor quantities are constructed from $q^{(0)}$ and
$Q^{(0)}$, and indices are raised and lowered by ${}^d g_{ij}$ and ${}^D g_{ab}$ in Eq. (\ref{eq:a1}), 
i.e.,
\begin{equation}
 \label{eq:aa2}
 \begin{split}
q^{(0)}_i &= -\frac{1}{ k_r^{(0)}} q^{(0)}_{|i}, \quad  q^{(0)}_{ij} = \frac{1}{ k_r^{(0)2}} 
q^{(0)}_{|ij} + \frac{1}{d} {}^d g_{ij} q^{(0)} , \\
Q^{(0)}_a &= -\frac{1}{ k_R^{(0)}} Q^{(0)}_{|a}, \quad  Q^{(0)}_{ab} = \frac{1}{ k_R^{(0)2}} 
Q^{(0)}_{|ab} + \frac{1}{D} {}^D g_{ab} Q^{(0)} .
\end{split}
\end{equation} 

The vector Helmholtz equations are 
\begin{equation}
 \label{eq:aa3}
{q^{(1)i}}^{|j}_{|j} + k_r^{(1)2} q^{(1)i} = 0, \quad
{Q^{(1)a}}^{|b}_{|b} + k_R^{(1)2} Q^{(1)a} = 0,
\end{equation} 
where $q^{(1)i}$ and $Q^{(1)a}$ are divergenceless, intrinsically vector quantities and 
do not reduce to any scalars. Tensor (constructed) quantities are
\begin{equation}
 \label{eq:aa4}
q^{(1)}_{ij} = - \frac{1}{2k_r^{(1)}} (q^{(1)}_{i|j} + q^{(1)}_{j|i}),  \quad
Q^{(1)}_{ab} = - \frac{1}{2k_R^{(1)}} (Q^{(1)}_{a|b} + Q^{(1)}_{b|a}).
\end{equation} 

The tensor Helmholtz equations are 
\begin{equation}
 \label{eq:aa5}
{q^{(2)ij}}^{|k}_{|k} + k_r^{(2)2} q^{(2)ij} = 0, \quad
{Q^{(2)ab}}^{|c}_{|c} + k_R^{(2)2} Q^{(2)ab} = 0,
\end{equation} 
where $q^{(2)ij}$ and $Q^{(2)ab}$ are divergenceless and traceless, intrinsically tensor quantities. 

\subsection{Gauge transformations}
Gauge transformations are shown in the following three modes.
\subsubsection{The scalar mode}
Coordinate transformations are expressed using three functions of $t$ as
\begin{equation}
 \label{eq:aa6}
 \begin{split}
\bar{t} &= t + T q^{(0)} Q^{(0)}, \\
\bar{x}^i &= x^i + l^{(0)} q^{(0)i} Q^{(0)}, \quad
\bar{X}^a = X^a + L^{(0)} q^{(0)} Q^{(0)a},
\end{split}
\end{equation} 
where $T, l,$ and $L$ are functions of $t$. For these transformations, the metric perturbations 
transform as
\begin{equation}
 \label{eq:aa7}
 \begin{split}
\bar{A} &= A - \dot{T}, \\
\bar{b}^{(0)} &= b^{(0)} +  (k_r^{(0)}/r) T + r \dot{l}^{(0)},  \\
\bar{B}^{(0)} &= B^{(0)} +  (k_R^{(0)}/R) T + R \dot{L}^{(0)},  \\
\bar{h}_L^{(0)} &= h_L^{(0)} -  (k_r^{(0)}/d) l^{(0)} - (\dot{r}/r) T ,  \\
\bar{H}_L^{(0)} &= H_L^{(0)} -  (k_R^{(0)}/D) L^{(0)} - (\dot{R}/R) T ,  \\
\bar{h}_T^{(0)} &= h_T^{(0)} + k_r^{(0)} l^{(0)}  ,  \\
\bar{H}_T^{(0)} &= H_T^{(0)} +  k_R^{(0)} L^{(0)}  ,  \\
\bar{G}^{(0)} &= G^{(0)} +k_R^{(0)} \frac{r}{2R} l^{(0)} +k_r^{(0)} \frac{R}{2r} L^{(0)}.
\end{split}
\end{equation} 
The fluid quantities transform as
\begin{equation}
 \label{eq:aa8}
 \begin{split}
\bar{v}^{(0)} &= v^{(0)} +  r \dot{l}^{(0)},  \quad
\bar{V}^{(0)} = V^{(0)} + R \dot{L}^{(0)},  \\
\bar{\delta} &= \delta  - \frac{\dot{\rho}}{\rho} T, \\
\bar{\pi}_L &= \pi_L  - \frac{\dot{p}}{p} T, \quad
\bar{\Pi}_L = \Pi_L  - \frac{\dot{p}}{p} T.
\end{split}
\end{equation} 

\subsubsection{The vector mode}
Coordinate transformations are expressed as
\begin{equation}
 \label{eq:aa9}
\bar{x}^i = x^i + l^{(1)} q^{(1)i} Q^{(0)}, \quad
\bar{X}^a = X^a + L^{(1)} q^{(0)} Q^{(1)a},
\end{equation} 
and the metric perturbations and fluid quantities transform as
\begin{equation}
 \label{eq:aa10}
 \begin{split}
\bar{b}^{(1)} &= b^{(1)} + r \dot{l}^{(1)},  \quad
\bar{B}^{(1)} = B^{(1)} + R \dot{L}^{(1)},  \\
\bar{h}_T^{(1)} &= h_T^{(1)} +  k_r^{(1)} l^{(1)}  ,  \quad
\bar{H}_T^{(1)} = H_T^{(1)} +  k_R^{(1)} L^{(1)}  ,  \\
\bar{G}^{(10)} &= G^{(10)} +k_R^{(0)} \frac{r}{2R} l^{(1)} ,\
\bar{G}^{(01)} = G^{(01)} +k_r^{(0)} \frac{R}{2r} L^{(1)} ,\\
\bar{G}^{(11)} &= G^{(11)} ,
\end{split}
\end{equation} 
and
\begin{equation}
 \label{eq:aa11}
\bar{v}^{(1)} = v^{(1)} +  r \dot{l}^{(1)},  \quad
\bar{V}^{(1)} = V^{(1)} + R \dot{L}^{(1)}. 
\end{equation} 

\subsubsection{The tensor mode}
All quantities are gauge-invariant themselves.

\section{Approximate solutions for $x (\equiv k_r^{(0)} \tau^{4/3}) \gg 1$ and $\mu/x \ll 1$}
\subsection{The case of $\omega = 1$}
For $\Phi_h = \Phi_{h0} (x) \exp{i\omega x}$ in Eq. (\ref{eq:b19}),
we have
\begin{equation}
 \label{eq:ab1}
\Phi_{h,x} = [i\omega \Phi_{h0} (x) + \Phi_{h0} (x)_{,x}] \exp{i \omega x}
\end{equation} 
and [the second term $ \Phi_{h0} (x)_{,x} /$ the first term $\omega \Phi_{h0} (x) $] 
is $\sim 1/x \ (\ll 1)$. When we take only the lowest-order terms with respect to $1/x$
 in  Eqs. (\ref{eq:b12}) -  (\ref{eq:b14}), we obtain  
\begin{equation}
 \label{eq:ab2}
 \begin{split}
\frac{1}{2}i x^{-1/2} \Phi_{G0} &= \Phi_{h0} + \Phi_{H0} -\frac{1}{3} x^{-3/4} \Phi_{60}, \\
(\frac{1}{9} -1) \Phi_{h0} &= \frac{4}{3} \Phi_{H0} - \frac{2}{9} x^{-3/4} \Phi_{60}, \\
(\frac{2}{3} -1) \Phi_{H0} &= -\frac{1}{9} \Phi_{h0} + \frac{1}{9} x^{-3/4} \Phi_{60} - 
\frac{i}{2} x^{-1/2} \Phi_{G0}.
\end{split}
\end{equation} 
From these three equations, we find 
\begin{equation}
 \label{eq:ab3}
\Phi_{H0} = -\frac{1}{3} \Phi_{h0}, \quad
\Phi_{60} = 2 x^{3/4} \Phi_{h0}, \quad
\Phi_{G0} = 0.
\end{equation} 
In order to derive the $x$-dependence of $\Phi_h$, we consider the next-order terms with 
respect to $1/x$. Put 
\begin{equation}
 \label{eq:ab4}
\Phi_{H0} = -\frac{1}{3} \Phi_{h0} + i \Delta \Phi_H, 
\end{equation} 
and 
\begin{equation}
 \label{eq:ab5}
\Phi_{G0} =  \Delta \Phi_G .
\end{equation} 
From Eq.  (\ref{eq:b12}), we obtain, taking account of the next-order terms,

\begin{equation}
 \label{eq:ab5a}
  \begin{split}
i \Phi_{60} + \Phi_{60,x} &= -\frac{3}{4} x^{-1} \Phi_{60} + 3x^{3/4} [i(\Phi_{h0}+\Phi_{H0})\\
 &+ \Phi{h0,x} +\Phi{H0,x} + (2\Phi_{h0} +\Phi_{H0})/x] + \frac{3}{2} x^{1/4} \Delta\Phi_G, 
  \end{split}
\end{equation} 
where $\Phi_{60,x}$ and $x^{-1}\Phi_{60}$ are of the next order. For them, we can use the relation
$\Phi_{h0} +\Phi_{H0} - \frac{1}{3} x^{-3/4} \Phi_{60} = 0$ to eliminate $\Phi_{60}$. Then we obtain
\begin{equation}
 \label{eq:ab7}
\Phi_{60} - 3x^{3/4} ( \Phi_{h0} + \Phi_{H0}) = -\frac{3}{2} i [(\Phi_{h0}-\Phi_{H0})
x^{-1/4} + \Delta \Phi_G x^{1/4} ]
\end{equation} 
which includes additional next-order terms in contrast to the first line of Eq. (\ref{eq:ab2}). 
Next from Eq.  (\ref{eq:b11}), we have
\begin{equation}
 \label{eq:ab7a}
2i \Delta \Phi_{G,x} + \frac{i}{x} \Delta \Phi_G = 2 x^{-1/2} (\Phi_{h0} + \Phi_{H0} - \frac{1}{3} x^{-3/4} \Phi_{60}). 
\end{equation} 
Substituting Eq.(\ref{eq:ab7}) into this equation, we obtain 
\begin{equation}
 \label{eq:ab6}
 \Delta \Phi_{G,x} =  \frac{1}{2} ( \Phi_{h0} - \Phi_{H0}) x^{-3/2} = \frac{2}{3}  
 \Phi_{h0} x^{-3/2},  
\end{equation} 
where we have used Eq. (\ref{eq:ab3}) to get the second equality in the next-order terms.

For the latter two lines of Eq. (\ref{eq:ab2}), we have the following additional next-order terms,
expressed as
\begin{equation}
 \label{eq:ab8}
(1 - \frac{1}{9}) \Phi_{h0} + \frac{4}{3} \Phi_{H0} -\frac{2}{9} x^{-3/4} \Phi_{60}  = 
-\frac{2}{9} [x^{-3/4} \Phi_{60} - 3(\Phi_{h0}+\Phi_{H0}) - 3i \Delta \Phi_H],
\end{equation} 
\begin{equation}
 \label{eq:ab9}
(1 - \frac{2}{3}) \Phi_{H0} + \frac{1}{9} \Phi_{h0} +\frac{1}{9} x^{-3/4} \Phi_{60}  = 
\frac{1}{9} [x^{-3/4} \Phi_{60} - 3(\Phi_{h0}+\Phi_{H0}) + 6i \Delta \Phi_H].
\end{equation} 

Taking Eqs.(\ref{eq:ab3}), (\ref{eq:ab7}), (\ref{eq:ab8}), and (\ref{eq:ab9}) into account, we obtain from Eqs. 
 (\ref{eq:b13}) and  (\ref{eq:b14})
\begin{equation}
 \label{eq:ab10}
2 \Phi_{h0,x} =  - \frac{1}{3x} \Phi_{h0} +\frac{2}{3} \Delta \Phi_H +
\frac{1}{3} x^{-1/2} \Delta \Phi_G,
\end{equation} 
\begin{equation}
 \label{eq:ab11}
-\frac{2}{3} \Phi_{h0,x} =  - \frac{1}{3x} \Phi_{h0} +\frac{2}{3} \Delta \Phi_H -
\frac{2}{3} x^{-1/2} \Delta \Phi_G,
\end{equation} 
respectively. If we eliminate $\Delta \Phi_H$ from these two equations, we obtain
\begin{equation}
 \label{eq:ab12}
\Phi_{h0,x} =   \frac{3}{8}  x^{-1/2} \Delta \Phi_G.
\end{equation} 
Using Eq. (\ref{eq:ab6}), we obtain
\begin{equation}
 \label{eq:ab13}
(\Phi_{h0,x} x^{1/2})_{,x} =   \frac{1}{4} \Phi_{h0}  x^{-3/2}. 
\end{equation} 
For $\Phi_{h0} \propto x^{\alpha}$, the equation for $\alpha$ leads to Eq. (\ref{eq:b24}) and
its solutions lead to Eq. (\ref{eq:b25}).

\subsection{The case of $\omega = 1/3$}
From lowest-order terms in Eqs. (\ref{eq:b12}) -  (\ref{eq:b14}), we obtain the relations $\Phi_{h0}
= \Phi_{H0}$ and $\Phi_{G0} = 0$, as shown in Eqs. (\ref{eq:b29}) and (\ref{eq:b22}).
Now, to derive the $x$-dependence of $\Phi_{h0}$, we consider the next-order terms
with respect to $1/x$, such as
\begin{equation}
 \label{eq:ab14}
\Phi_{H0} - \Phi_{h0} = \Delta \Phi_H,
\end{equation} 
and 
\begin{equation}
 \label{eq:ab15}
\Phi_{G0} = \Delta \Phi_G.
\end{equation} 
From Eqs. (\ref{eq:b13}) and  (\ref{eq:b14}), we obtain 
\begin{equation}
 \label{eq:ab16}
\Phi_{h0,x} = \frac{1}{6x} \Phi_{h0} + \frac{5}{3x}  \Phi_{H0} - i (2\Phi_{H0} 
-\frac{1}{3} x^{-3/4} \Phi_{60}), 
\end{equation} 
\begin{equation}
 \label{eq:ab17}
\Phi_{H0,x} = -\frac{7}{12x} \Phi_{h0} - \frac{19}{12x}  \Phi_{H0} + i\frac{1}{6} (5\Phi_{H0} 
+\Phi_{h0} -x^{-3/4} \Phi_{60}) - \frac{1}{4} x^{-1/2} \Delta \Phi_G.
\end{equation} 

From Eq. (\ref{eq:b12}),  on the other hand, we have
\begin{equation}
 \label{eq:ab18}
\Phi_{60} - 3 x^{3/4} (\Phi_{h0} + \Phi_{H0}) = - \frac{3}{2} i (x^{-1/4} \Delta \Phi_H + 
x^{1/4} \Delta \Phi_G).
\end{equation} 
From the consistency of Eq. (\ref{eq:b11}), moreover, we obtain
\begin{equation}
 \label{eq:ab19}
\Delta \Phi_G = 0.
\end{equation} 

Using Eq. (\ref{eq:ab18}), Eqs.  (\ref{eq:ab16}) and (\ref{eq:ab17}) lead to
\begin{equation}
 \label{eq:ab20}
\Phi_{h0,x} = \frac{11}{6x} \Phi_{h0} - i \Delta \Phi_H,
\end{equation} 
\begin{equation}
 \label{eq:ab21}
\Phi_{h0,x} = - \frac{13}{6x} \Phi_{h0} +\frac{1}{3} i \Delta \Phi_H,
\end{equation} 
where higher-order terms $\sim \Delta \Phi_H/x$ were neglected.
From these two equations, we obtain
\begin{equation}
 \label{eq:ab22}
\Phi_{h0,x} = - \frac{7}{6x} \Phi_{h0} ,
\end{equation} 
or $\Phi_{h0} \propto x^{-7/6}$.  For $\Delta \Phi_H$, we have
\begin{equation}
 \label{eq:ab23}
\Delta \Phi_H = - (3i/x) \Phi_{h0}.
\end{equation} 

\subsection{The power-type solutions}
When we use Eq. (\ref{eq:b31}) for the expressions of $\Phi_h, \Phi_H, \Phi_6,$ and $\Phi_G$, 
we obtain from Eqs. (\ref{eq:b11}) - (\ref{eq:b14})
\begin{equation}
 \label{eq:ab24}
 \begin{split}
[x^{\alpha_G} (1 - \frac{1}{4} x^{-2}) &+ (\alpha_G)^2 x^{\alpha_G -2}] \Phi_{G0} = 
2 x^{\alpha_h -1/2} \Phi_{h0} \\
&+ 2 x^{\alpha_H -1/2} \Phi_{H0} - \frac{2}{3} x^{\alpha_6 -5/4} \Phi_{60},
\end{split}
\end{equation} 
\begin{equation}
 \label{eq:ab25}
 \begin{split}
(\alpha_6 + \frac{3}{4}) x^{\alpha_6 -1} \Phi_{60} &= 3[(\alpha_h +2) x^{\alpha_h -1/4} \Phi_{h0}\\
&+ (\alpha_H +1) x^{\alpha_H -1/4} \Phi_{H0}] + \frac{3}{2} x^{\alpha_G +1/4} \Phi_{G0},
\end{split}
\end{equation} 
\begin{equation}
 \label{eq:ab26}
 \begin{split}
\{\frac{1}{9} x^{\alpha_h} &+ [\alpha_h (\alpha_h - \frac{4}{3}) - \frac{20}{3}] x^{\alpha_h -2}\}
 \Phi_{h0} = [\frac{4}{3} x^{\alpha_H} 
+ \frac{10}{3} (\alpha_H +1) x^{\alpha_H -2}] \Phi_{H0}\\
 &-\frac{2}{9} x^{\alpha_6 -3/4} 
(1 + 6x^{-2}) \Phi_{60} + \frac{23}{12} x^{\alpha_G -3/2}  \Phi_{G0} ,
\end{split}
\end{equation} 
\begin{equation}
 \label{eq:ab27}
 \begin{split}
\{\frac{2}{3} x^{\alpha_H} &+ [\alpha_H (\alpha_H + \frac{13}{6}) + \frac{7}{6}] x^{\alpha_H -2}\} \Phi_{H0} 
= -[\frac{1}{9} x^{\alpha_h} + \frac{7}{6} (\alpha_h +2) x^{\alpha_h -2}] \Phi_{h0} \\
&+\frac{1}{9} x^{\alpha_6 -3/4} (1 + \frac{33}{4} x^{-2}) \Phi_{60} 
- \frac{1}{2} (\alpha_G +\frac{5}{3}) x^{\alpha_G -3/2}  \Phi_{G0} .
\end{split}
\end{equation} 

As the consistent sets of powers $(\alpha_h, \alpha_H, \alpha_6, \alpha_G)$ and 
constants $(\Phi_{h0} ,
\Phi_{H0}, \Phi_{60}, \Phi_{G0} )$ for these equations, it is found that
\begin{equation}
 \label{eq:ab28}
\alpha_h = \alpha_H =  \alpha_6 - 3/4 = \alpha_G + 1/2,
\end{equation} 
and
\begin{equation}
 \label{eq:ab29}
\frac{1}{9} \Phi_{h0} =\frac{4}{3} \Phi_{H0} -\frac{2}{9} \Phi_{60},
\end{equation} 
\begin{equation}
 \label{eq:ab30}
\frac{2}{3} \Phi_{H0} = -\frac{1}{9} \Phi_{h0} + \frac{1}{9} \Phi_{60},
\end{equation} 
\begin{equation}
 \label{eq:ab31}
(\alpha_6 +\frac{3}{4}) \Phi_{60} = 3[(\alpha_h +2) \Phi_{h0} + (\alpha_H +1) \Phi_{H0}]
+ \frac{3}{2} \Phi_{G0},
\end{equation} 
\begin{equation}
 \label{eq:ab32}
\frac{1}{2} \Phi_{G0} = \Phi_{h0} + \Phi_{H0} - \frac{1}{3} \Phi_{60},
\end{equation} 
by taking the lowest-order terms with respect to $1/x$ in Eqs. (\ref{eq:ab24}) - (\ref{eq:ab27}).

From Eqs. (\ref{eq:ab29}) and (\ref{eq:ab30}), we obtain
\begin{equation}
 \label{eq:ab33}
\Phi_{h0} = 0 \quad {\rm and}  \quad  \Phi_{60} = 6 \Phi_{H0}.
\end{equation} 

Inserting Eq. (\ref{eq:ab33}) into Eqs. (\ref{eq:ab31}) and (\ref{eq:ab32}), we obtain
\begin{equation}
 \label{eq:ab34}
\Phi_{G0} = -2 \Phi_{H0} \quad {\rm and}  \quad  \alpha_6 = \frac{1}{2} \alpha_H - \frac{3}{4}.
\end{equation} 
From Eqs. (\ref{eq:ab28}),  (\ref{eq:ab33}), and (\ref{eq:ab34}), we have finally  Eqs. (\ref{eq:b32})
and (\ref{eq:b33}).

\section{Approximate solutions for $y (\equiv k_R^{(0)} \tau^{2/3}) \gg 1$ and $\nu y^2 \ll 1$}

\subsection{The case of $\omega = 1$}
For $\Phi_h = \Phi_{h0} (y) \exp{i\omega y}$ in Eq. (\ref{eq:b43}),
we have
\begin{equation}
 \label{eq:ac1}
\Phi_{h,y} = [i\omega \Phi_{h0} (y) + \Phi_{h0} (y)_{,y}] \exp{i \omega y}
\end{equation} 
and [the second term $ \Phi_{h0} (y)_{,y}/$ the first term $\omega \Phi_{h0} (y) $] is $\sim 1/y
\ (\ll 1)$. When we take only the lowest-order terms with respect to $1/y$ in  Eqs. (\ref{eq:b36}) - 
 (\ref{eq:b38}), we obtain  
\begin{equation}
 \label{eq:ac2}
 \begin{split}
 \Phi_{h0} &+ \Phi_{H0} -\frac{1}{3} y^{-3/2} \Phi_{60} + i y^{-2} \Phi_{G0}  = 0, \\
\Phi_{h0} &+ \frac{1}{3} \Phi_{H0} - \frac{2}{3} x^{-3/2} \Phi_{60} - 3 i y^{-2} \Phi_{G0} = 0, \\
 \Phi_{h0} &+ \frac{8}{3} \Phi_{H0} - \frac{1}{3} y^{-3/2} \Phi_{60} = 0.
\end{split}
\end{equation} 
From these three equations, we find 
\begin{equation}
 \label{eq:ac3}
\Phi_{H0} = -\frac{1}{10} \Phi_{h0}, \quad
\Phi_{60} =  \frac{11}{5} y^{3/2} \Phi_{h0}, \quad
\Phi_{G0} = \frac{i}{6} y^2 \Phi_{h0}.
\end{equation} 
Next, from  Eq. (\ref{eq:b35}), we have
\begin{equation}
2i \Phi_{G0,y} + i\Phi_{G0}/y = 2y (\Phi_{h0} + \Phi_{H0} - \frac{1}{3} y^{-3/2} \Phi_{60}),
\end{equation} 
and using the first line of Eq. (\ref{eq:ac2}), we obtain 
\begin{equation}
 \label{eq:ac4}
 \Phi_{G0,y}  = - \frac{3}{2} y^{-1} \Phi_{G0} ,
\end{equation} 
or $\Phi_{G0} \propto y^{-3/2}$. From Eq.  (\ref{eq:ac3}), therefore, we have
\begin{equation}
\Phi_{h0} \propto \Phi_{H0} \propto y^{-7/2}, \quad  \Phi_{60} \propto y^{-2}.
\end{equation} 

\subsection{The case of $\omega = 2/3$ and $\Phi_{G0} = 0$}
From the lowest-order terms in Eqs. (\ref{eq:b36}) -  (\ref{eq:b38}) with respect to $1/y$, 
we have in this case
\begin{equation}
 \label{eq:ac5}
\Phi_{H0} = \frac{2}{5} \Phi_{h0}, 
\end{equation} 
\begin{equation}
 \label{eq:ac6}
\Phi_{60} = \frac{2}{5} y^{3/2} \Phi_{h0}.
\end{equation} 
To derive the $y$-dependence of $\Phi_{h0}$, we consider the next-order terms
($\Delta \Phi_H$ and $\Delta \Phi_G$) with respect to $1/y$,  defined by
\begin{equation}
 \label{eq:ac7}
\Phi_{H0} = \frac{2}{5} \Phi_{h0} + i \Delta \Phi_H,
\end{equation} 
and 
\begin{equation}
 \label{eq:ac8}
\Phi_{G0} = \Delta \Phi_G.
\end{equation} 
From Eq. (\ref{eq:b35}), we obtain 
\begin{equation}
 \label{eq:ac9}
\Phi_{h0}  + \Phi_{H0} -\frac{1}{3} y^{-3/2} \Phi_{60} = \frac{i}{3y}(2\Delta \Phi_{G,y}
 + \frac{1}{y}\Delta \Phi_{G}) ,
\end{equation} 
and, from Eq. (\ref{eq:b36}),
\begin{equation}
 \label{eq:ac10}
 \begin{split} 
\frac{2}{3} i [\Phi_{60} &- 3y^{3/2} (\Phi_{h0} +\Phi_{H0})] = \frac{1}{y} [-\frac{3}{2}\Phi_{60}\\
&+ 3y^{3/2} (4 \Phi_{h0} + 2 \Phi_{H0})] - 3 y^{-1/2} \Delta \Phi_{G} .
 \end{split}
\end{equation} 
From Eqs. (\ref{eq:ac9}) and (\ref{eq:ac10}), we obtain
\begin{equation}
 \label{eq:ac11}
\Delta \Phi_{G,y} + \frac{11}{4}y^{-1} \Delta \Phi_G = \frac{243}{40} \Phi_{h0},
\end{equation} 
where higher-order terms $\sim \Delta \Phi_H/y$ and $\Delta \Phi_G/y$  were neglected and
Eq.(\ref{eq:ac5}) was used.

 Next, from  Eq. (\ref{eq:b37}), we have
\begin{equation}
 i [4 \Phi_{h0,y} - \frac{10}{3y}\Phi_{h0} - \frac{40}{3y} \Phi_{H0})] 
 - \frac{2}{3}y^{-3/2} \Phi_{60} 
 + \frac{8}{3}\Phi_{h0} + \frac{1}{3}\Phi_{H0} - \frac{2i}{y^2} \Delta \Phi_{G} = 0.
\end{equation} 
Using Eqs. (\ref{eq:ac5}),  (\ref{eq:ac7}), and  (\ref{eq:ac9}), this equation reduces to
\begin{equation}
 \label{eq:ac12}
\Phi_{h0,y} - \frac{13}{6y} \Phi_{h0} + \frac{1}{3y} (\Delta \Phi_{G,y} - \frac{1}{y} \Delta \Phi_G)
- \frac{5}{12} \Delta \Phi_H = 0.
\end{equation} 
From Eqs.  (\ref{eq:b38}) and  (\ref{eq:ac9}), moreover, we obtain
\begin{equation}
 \label{eq:ac13}
\Phi_{h0,y} + \frac{67}{12y} \Phi_{h0} - \frac{5}{24y} (2\Delta \Phi_{G,y} 
+ \frac{1}{y} \Delta \Phi_G)  = 0.
\end{equation} 
Here let us assume the following functional forms for $\Delta \Phi_G$ and $\Phi_{h0}$
\begin{equation}
 \label{eq:ac14}
\Delta \Phi_G = \Phi_{G10} \ y^{\alpha +1} \quad {\rm and} \quad \Phi_{h0} = 
\Phi_{h01} \ y^\alpha,
\end{equation} 
where $\Phi_{G10}, \Phi_{h01}, $ and $\alpha$ are constants.
Then from Eqs. (\ref{eq:ac11}),  (\ref{eq:ac12}), and  (\ref{eq:ac13}) , we obtain
\begin{equation}
 \label{eq:ac15}
(2\alpha + \frac{15}{2})  \Delta \Phi_{G10} = \frac{243}{20} \Phi_{h01}, 
\end{equation} 
\begin{equation}
 \label{eq:ac16}
(\alpha - \frac{13}{6})  \Delta \Phi_{h01} = -\frac{1}{3} \alpha \Phi_{G10} + \frac{5}{12} \Phi_{H0}, 
\end{equation} 
\begin{equation}
 \label{eq:ac17}
(\alpha + \frac{67}{12})  \Delta \Phi_{h01} = \frac{5}{24} (2\alpha +3) \Phi_{G10} . 
\end{equation} 
From Eqs.  (\ref{eq:ac15}) and   (\ref{eq:ac17}), it is found that $\alpha$ satisfies 
Eq.  (\ref{eq:b52}) , and its solution is expressed as Eq.  (\ref{eq:b53}).

\subsection{The case of $\omega = (2/3)^{1/2}$ and $\Phi_{G0} = 0$}
From the lowest-order terms in Eqs. (\ref{eq:b36}) -  (\ref{eq:b38}), we have in this case
\begin{equation}
 \label{eq:cc1}
\Phi_{H0} = 0, 
\end{equation} 
\begin{equation}
 \label{eq:cc2}
\Phi_{60} = 3 y^{3/2} \Phi_{h0}.
\end{equation} 
To derive the $y$-dependence of $\Phi_{h0}$, we consider the next-order terms
($\Delta \Phi_H$ and $\Delta \Phi_G$) with respect to $1/y$,  defined by
\begin{equation}
 \label{eq:cc3}
\Phi_{H0} = i \Delta \Phi_H,
\end{equation} 
and 
\begin{equation}
 \label{eq:cc4}
\Phi_{G0} = \Delta \Phi_G.
\end{equation} 
From Eq. (\ref{eq:b35}), we obtain 
\begin{equation}
 \label{eq:cc5}
\Phi_{h0}  + \Phi_{H0} -\frac{1}{3} y^{-3/2} \Phi_{60} = \frac{i\omega}{2y}(2\Delta \Phi_{G,y}
 + \frac{1}{y}\Delta \Phi_{G}) ,
\end{equation} 
and from Eq. (\ref{eq:b36})
\begin{equation}
 \label{eq:cc6}
 \begin{split} 
 i \omega [\Phi_{60} &- 3y^{3/2} (\Phi_{h0} +\Phi_{H0})] = \frac{1}{y} [-\frac{3}{2}\Phi_{60}\\
&+ 3y^{3/2} (4 \Phi_{h0} + 2 \Phi_{H0})] - 3 y^{-1/2} \Delta \Phi_{G}.
\end{split}
\end{equation} 
From Eqs. (\ref{eq:cc5}) and (\ref{eq:cc6}), we obtain
\begin{equation}
 \label{eq:cc7}
\Delta \Phi_{G,y} + \frac{2}{y} \Delta \Phi_G = \frac{15}{4} \Phi_{h0},
\end{equation} 
where higher-order terms $\sim \Delta \Phi_H/y$ and $\Delta \Phi_G/y$  were neglected.

 Next, from  Eq. (\ref{eq:b37}), we have
\begin{equation}
 \label{eq:cc8}
 i \omega[6 \Phi_{h0,y} - \frac{5}{y}\Phi_{h0} - \frac{20}{y} \Phi_{H0}] 
 - \frac{2}{3}y^{-3/2} \Phi_{60} 
 + (4 -3\omega^2)\Phi_{h0} + \frac{1}{3}\Phi_{H0} - \frac{3i\omega}{y^2} \Delta \Phi_{G} = 0 .
\end{equation} 
Using Eqs. (\ref{eq:cc1}),  (\ref{eq:cc3}), and  (\ref{eq:cc5}), this equation reduces to
\begin{equation}
 \label{eq:cc9}
\Phi_{h0,y} - \frac{5}{6y} \Phi_{h0} + \frac{1}{3y} (\Delta \Phi_{G,y} - \frac{1}{y} \Delta \Phi_G)
- \frac{5}{18\omega} \Delta \Phi_H = 0.
\end{equation} 
From Eqs.  (\ref{eq:b38}) and  (\ref{eq:cc5}), moreover, we obtain
\begin{equation}
 \label{eq:cc10}
\frac{7}{y} \Phi_{h0} - \frac{1}{y} (\Delta \Phi_{G,y} 
+ \frac{1}{2y} \Delta \Phi_G) -\frac{2}{3\omega} \Delta \Phi_H = 0.
\end{equation} 
Eliminating $\Delta \Phi_H$ from Eqs. (\ref{eq:cc9}) and (\ref{eq:cc10}),
we obtain
\begin{equation}
 \label{eq:cc11}
\Phi_{h0,y} -\frac{15}{4y} \Phi_{h0} - \frac{3}{4y} (\Delta \Phi_{G,y} 
- \frac{1}{6y} \Delta \Phi_G) = 0.
\end{equation} 
Here let us assume the following functional forms for $\Delta \Phi_H, \Delta \Phi_G$ and
 $\Phi_{h0}$
\begin{equation}
 \label{eq:cc12}
\Delta \Phi_H = \Phi_{H01}\ y^{\beta-1}, \ \Delta \Phi_G = \Phi_{G10} \ y^{\beta +1} ,
\quad {\rm and} \quad \Phi_{h0} = \Phi_{h01} \ y^\beta,
\end{equation} 
where $\Phi_{H01}, \Phi_{G10}, \Phi_{h01} $ and $\beta$ are constants.
Then from Eqs. (\ref{eq:cc7}),  (\ref{eq:cc9}), and  (\ref{eq:cc11}) , we obtain
\begin{equation}
 \label{eq:cc13}
(\beta + 3)   \Phi_{G10} = \frac{15}{4} \Phi_{h01}, 
\end{equation} 
\begin{equation}
 \label{eq:cc14}
(\beta - \frac{5}{6})   \Phi_{h01} = -\frac{1}{3} \beta \Phi_{G10} + \frac{5}{18\omega} \Phi_{H01}, 
\end{equation} 
\begin{equation}
 \label{eq:cc15}
(\beta - \frac{15}{4})  \Phi_{h01} = -\frac{3}{4} (\beta +\frac{5}{6}) \Phi_{G10} . 
\end{equation} 
From Eqs.  (\ref{eq:cc13}) and   (\ref{eq:cc15}), it is found that $\beta$ satisfies 
Eq.  (\ref{eq:cc16}) , and its solution is expressed as Eq.  (\ref{eq:cc17}).
Moreover, $\Phi_{H01}$ and $\Phi_{G10}$ are expressed in terms of $\Phi_{h01}$ using 
Eqs. (\ref{eq:cc13}) and  (\ref{eq:cc14}).

\subsection{The power-type solutions}
When we use Eq. (\ref{eq:b54}) for the expressions of $\Phi_h, \Phi_H, \Phi_6,$ and $\Phi_G$, 
we obtain, from Eqs. (\ref{eq:b35}) - (\ref{eq:b38}),
\begin{equation}
 \label{eq:ac18}
 \begin{split}
[y^{\alpha_G} (1 -  y^{-2}) &+ (\alpha_G)^2 y^{\alpha_G -2}] \Phi_{G0} = 
2 y^{\alpha_h +1} \Phi_{h0} \\
&+ 2 y^{\alpha_H +1} \Phi_{H0} - \frac{2}{3} y^{\alpha_6 - 1/2} \Phi_{60},
\end{split}
\end{equation} 
\begin{equation}
 \label{eq:ac19}
 \begin{split}
(\alpha_6 + \frac{3}{2}) y^{\alpha_6 -1} \Phi_{60} &= 3y^{3/2} [(\alpha_h +4) y^{\alpha_h -1} \Phi_{h0}\\
&+ (\alpha_H +2) y^{\alpha_H -1} \Phi_{H0}] - 3 y^{\alpha_G -1/2} \Phi_{G0},
\end{split}
\end{equation} 
\begin{equation}
 \label{eq:ac20}
 \begin{split}
\{4 y^{\alpha_h} &+ [3 \alpha_h (\alpha_h - 1) - 5\alpha_h -80] y^{\alpha_h -2}\} \Phi_{h0} 
+ [\frac{1}{3} y^{\alpha_H} 
-  20 (\alpha_H +2) x^{\alpha_H -2}] \Phi_{H0}\\ &-\frac{2}{3} (1- 4y^{-2}) y^{\alpha_6 -3/2} 
\Phi_{60} + (23 -3\alpha_G) y^{\alpha_G -3}  \Phi_{G0}  = 0,
\end{split}
\end{equation} 
\begin{equation}
 \label{eq:ac21}
 \begin{split}
\{\frac{1}{3} y^{\alpha_H} &+ [3\alpha_H (\alpha_H -1) + 16\alpha_H +14] y^{\alpha_H -2}\} \Phi_{H0}  + [-y^{\alpha_h} 
+  (7 \alpha_h +28) y^{\alpha_h -2}] \Phi_{h0}\\
&+ \frac{1}{3} (1- 33y^{-2}) y^{\alpha_6 -3/2} 
 \Phi_{60} - 7 y^{\alpha_G -3}  \Phi_{G0} = 0 .
\end{split}
\end{equation} 

As the consistent sets of powers $(\alpha_h, \alpha_H, \alpha_6, \alpha_G)$ and 
constants $(\Phi_{h0} ,
\Phi_{H0}, \Phi_{60}, \Phi_{G0} )$ for these equations, it is found that
\begin{equation}
 \label{eq:ac22}
\alpha_h = \alpha_H =  \alpha_6 - 3/2 = \alpha_G -1,
\end{equation} 
and
\begin{equation}
 \label{eq:ac23}
4 \Phi_{h0} + \frac{1}{3} \Phi_{H0} - \frac{2}{3} \Phi_{60} = 0,
\end{equation} 
\begin{equation}
 \label{eq:ac24}
\frac{1}{3} \Phi_{H0}  - \Phi_{h0} + \frac{1}{3} \Phi_{60} = 0,
\end{equation} 
\begin{equation}
 \label{eq:ac25}
(\alpha_6 +\frac{3}{2}) \Phi_{60} = 3[(\alpha_h +4) \Phi_{h0} + (\alpha_H +2) \Phi_{H0}]
-3 \Phi_{G0},
\end{equation} 
\begin{equation}
 \label{eq:ac26}
\frac{1}{2} \Phi_{G0} = \Phi_{h0} + \Phi_{H0} - \frac{1}{3} \Phi_{60},
\end{equation} 
by taking the lowest-order terms with respect to $1/y$ in Eqs. (\ref{eq:ac18}) - (\ref{eq:ac21}).

From Eqs. (\ref{eq:ac23}) and (\ref{eq:ac24}), we obtain
\begin{equation}
 \label{eq:ac27}
\Phi_{H0} = -2 \Phi_{h0},
\end{equation} 
Inserting this into Eq. (\ref{eq:ac24}),
\begin{equation}
 \label{eq:ac28}
\Phi_{60} = 5 \Phi_{h0}, 
\end{equation} 
and from Eq. (\ref{eq:ac26}), moreover,
\begin{equation}
 \label{eq:ac29}
\Phi_{G0} = -\frac{16}{3} \Phi_{h0}.
\end{equation} 
For powers, we obtain, from Eq.  (\ref{eq:ac25}),
\begin{equation}
 \label{eq:ac30}
5(\alpha_6 + 3/2) = 3 \alpha_h - 6 \alpha_H + 16,
\end{equation} 
and, using Eq.(\ref{eq:ac22}),
\begin{equation}
 \label{eq:ac31}
\alpha_h = \alpha_H = 1/8, \quad \alpha_6 = 13/8, \quad {\rm and} \quad 
\alpha_G = 9/8.
\end{equation} 
%


\end{document}